\documentclass[iop]{emulateapj}

\slugcomment{to appear in ApJ}

\shorttitle{The Search for Young Planets}
\shortauthors{Bailey et al.}

\begin{document}


\title{Precise Infrared Radial Velocities from Keck/NIRSPEC 
and the Search for Young Planets}


\author{John I. Bailey, III\altaffilmark{1}}
\affil{University of Alabama Huntsville, Physics Department,
301 Sparkman Drive, 201B Optics Building, Huntsville, AL 35899}
\email{baileyji@umich.edu}

\author{Russel J. White}
\affil{Georgia State University, Department of Physics and 
Astronomy, 29 Peachtree Center Avenue, Science Annex, Suite 400, Atlanta, 
GA 30303}
\email{white@chara.gsu.edu}

\author{Cullen H. Blake}
\affil{Department of Astrophysical Sciences, Princeton University, 
Peyton Hall, Ivy Lane, Princeton, NJ 08544}

\author{Dave Charbonneau}
\affil{Harvard-Smithsonian Center for Astrophysics, 60 Garden 
Street, Cambridge, MA 02138}

\author{Travis S. Barman}
\affil{Lowell Observatory, 1400 West Mars Hill Road, Flagstaff,
AZ 86001}

\author{Angelle M. Tanner}
\affil{Georgia State University, Department of Physics and 
Astronomy, 29 Peachtree Center Avenue, Science Annex, Suite 400, Atlanta, 
GA 30303}

\and
\author{Guillermo Torres}
\affil{Harvard-Smithsonian Center for Astrophysics, 60 Garden 
Street, Cambridge, MA 02138}

\altaffiltext{1}{Current Address: Department of Astronomy, University 
of Michigan, 830 Dennison Building, 500 Church Street, Ann Arbor, MI 
48109}

\altaffiltext{6}{Current Address: Mississippi State University, Department of Physics and 
Astronomy, 355 Lee Boulevard, Hilbun Hall, Starkville, 
MS 39762}




\begin{abstract}
We present a high-precision infrared radial velocity study of late-type 
stars using spectra obtained with NIRSPEC at the W. M. Keck Observatory.  
Radial velocity precisions of $50$ m/s are achieved for old field 
mid-M dwarfs using telluric features for precise wavelength calibration.  
Using this technique, 20 young stars in the $\beta$ Pic (age $\sim 12$ 
Myr) and TW Hya (age $\sim 8$ Myr) Associations were monitored over
several years to search for low mass companions; we also included the
chromospherically active field star GJ 873 (EV Lac) in this survey.
Based on comparisons with previous optical 
observations of these young active stars, radial velocity measurements at 
infrared wavelengths mitigate the radial velocity noise caused by star 
spots by a factor of $\sim 3$.  Nevertheless, star spot noise is still 
the dominant source of measurement error for young stars at 2.3 $\mu$m, 
and limits the precision to $\sim 77$ m/s for the slowest rotating
stars ($v$sin$i$ $<$ 6 km/s), increasing to $\sim 168$ m/s for
rapidly rotating stars ($v$sin$i$ $> 12$ km/s).
The observations reveal both GJ 3305 
and TWA 23 to be single-lined spectroscopic binaries; in the case of GJ 
3305, the motion is likely caused by its 0\farcs09 companion, identified 
after this survey began.  The large amplitude, short-timescale variations 
of TWA 13A are indicative of a hot Jupiter-like companion, but the available 
data are insufficient to confirm this.  We label it as a candidate 
radial velocity variable.  
For the remainder of the sample, these observations exclude the presence of 
any 'hot' (P $< 3$ days) companions more massive than 8 M$_{Jup}$, and
any 'warm' (P $< 30$ day) companions more massive than 17 M$_{Jup}$, on
average.  Assuming an edge-on orbit for the edge-on disk system AU Mic,
these observations exclude the presence of any hot Jupiters more massive
than 1.8 M$_{Jup}$ or warm Jupiters more massive than 3.9 M$_{Jup}$.
\end{abstract}


\keywords{stars: pre-main sequence --- starspots --- 
planets and satellites: formation --- techniques: radial velocities ---
techniques: spectroscopic --- methods: data analysis}

\section{Introduction}

The remarkable discovery of extrasolar planets orbiting sun-like stars in 
1995 was accomplished by utilizing novel techniques that improved the 
radial velocity (RV) precision by two orders of magnitude (from $\sim$1 km/s to 
10 m/s), which was at last sufficient to search for the slight 
gravitational wobble induced by an orbiting gas giant planet upon its 
host star (Walker et al. 1995;  Mayor \& Queloz 1995; Marcy \& Butler 1996;
Cochran et al. 1997; but cf. Latham et al. 1989).
Since then an extensive international effort has begun 
to determine the ubiquity and basic properties of these initially 
elusive objects. Based on a modest extrapolation of current findings, we 
now know that roughly 12\% of Sun-like stars (F,G, and K spectral type) 
harbor gas giant planets within 30 AU \citep[e.g.][]{butler06, udrysantos07,
cumming08, wright10, howard10}.  While the masses of these planets are 
similar to the gas giants 
in our solar system (i.e. Jupiter and Saturn), their orbital properties are 
in many ways radically different.  Some orbit their host star in just a few 
days time (the ``hot Jupiters''), while others have highly eccentric orbits.
Planet-like companions have even been directly imaged at separations of 
more than 100 AU from their host star \citep{kalas08}.  

The dramatically 
different properties of extrasolar planets compared to those in our solar 
system forced theorists to reconsider the standard paradigm for gas giant 
planet formation. Now, planet formation scenarios generally 
fall into 2 categories, core accretion \citep{mizuno80} and disk instabilities 
\citep{boss97}.  Core accretion is the favored mechanism for producing giant 
planets with large core masses \citep[e.g.][]{sato05} and can naturally 
explain the enhanced frequency of planets around metal rich stars 
\citep[e.g.][]{robinson06}.  However, at wide separations ($\gtrsim$ 
10 AU) the timescale for planet formation via core accretion 
\citep{pollack96, alibert05} is longer than the typical disk dissipation 
timescale \citep[$\lesssim$ 10 Myr;][]{haisch01, briceno01, carpenter06, 
pascucci06}; it is thus an unlikely formation mechanism for the planets in 
very wide orbits about Fomalhaut and HR 8799 \citep{marois08, kalas08}.  
Disk instabilities, on the other hand, can form planets much more quickly 
($\lesssim$ 1 Myr), well before circumstellar disks are dispersed 
\citep[][but 
cf. Stamatellos \& Whitworth 2008; Meru \& Bate 2011]{boss97, boss04, 
mayer02, merubate10}.  This mechanism also works nearly as efficiently 
around low mass stars, with sufficiently massive disks, and seems a more 
plausible scenario for forming the known massive planets orbiting M dwarfs 
\citep{boss06}.  Of course, it is possible that both scenarios operate 
or dominate within different regimes of stellar and/or disk mass.

In addition to the uncertainties surrounding planet formation, it is also 
unclear how the orbits of newly formed planets dynamically evolve.  Since 
giant planets 
are believed to form beyond the snowline \citep[e.g.][]{kennedykenyon08},
inward orbital migration must occur in many systems to explain the wealth 
of short period planets; roughly 10\% of gas giant planets have 
separations less than 0.1 AU \citep{butler06}.  Interactions with disk 
material is widely believed to be the most efficient mechanism for inward 
migration \citep[e.g.][]{goldreichtremaine80, lin96}.  If most gas giants
migrate via this process, the hot Jupiters must become 'hot' within 
$\sim 10$ Myr after formation, prior to disk dissipation.  However, planet-disk 
interactions typically circularize orbits and maintain spin-orbit alignment 
with the star, so other processes such as dynamical scattering with other 
planets \citep{adamslaughlin03, raymond11} or a nearby star 
\citep{malmberg07, malmbergdavies09} may be needed to produce 
the broad extrasolar planet eccentricity distribution and the spin-orbit 
misaglignments of many transiting planets \citep[e.g.][]{pont09}.  These
dynamical interactions could take many hundreds of millions of years
\citep[e.g. see review by][]{lubowida10}.

One direct way to begin constraining the mechanisms and timescales of 
planet formation and migration is to search for planets around stars 
with ages of $\lesssim 10$ Myr, or essentially immediately after planets
could have formed.  Direct imaging searches for planets have
been more successful at this recently than RV searches.  Several 
massive Jupiter-like companions have been directly imaged around stars 
with these ages, though at somewhat wide separations \citep[10 to many 
100s of AU; e.g.][]{neuhauser05, lagrange10, ireland11}; these 
important discoveries are helping to motivate future high-contrast
imaging facilities and the science achievable with them
\citep[e.g.][]{macintosh08, katariasimon10}.
In many cases, however, planetary status relies upon comparisons 
with poorly tested evolutionary models, and the wide separations offer 
little constraint on the migration mechanism responsible for forming hot 
Jupiter-like planets.

Planets identified via RV variations, on the other hand, yield immediate 
mass lower limits ($m$sin$i$) and preferentially find massive short-period 
planets which most likely have migrated.  However, two universal properties 
of young stars inhibit measuring their RV precisely at optical 
wavelengths, and consequently they are often excluded in high precision 
RV surveys for planets.  The first of these is that young Sun-like stars 
are much cooler while young, and thus fainter at optical wavelengths.  
A 1 solar mass star, for example, has a spectral type of $\sim$ K5 at 
10 Myr \citep{baraffe98, siess00}.  This, coupled with the fact that 
all star forming regions are further away than 100 pc means that 
most young stars are simply too faint for high-precision RV work at optical 
wavelengths.  The second of these is that young stars are often modestly 
to rapidly rotating.  This rotation broadens stellar absorption features 
and, even more perniciously, generates chromospheric activity and a 
spotted stellar surface.  Several pioneering RV studies of young spotted 
stars have shown that their presence can induce low amplitude, periodic 
RV shifts, and thus mimic the effect of one or more orbiting planets 
\citep{queloz01, paulsonyelda06, huerta08, prato08, reiners10, crockett11, 
mahmud11}; amplitudes of 200 - 
700 m/s have been observed.  In a couple cases, Jupiter-like companions
have been reported orbiting young stars based on low-amplitude ($< 1$ 
km/s) RV variations \citep{setiawan08, hernanobispo10}, but subsequent 
observations indicate that these too were misidentifications caused by 
star spots \citep{huelamo08, figueira10}.

One promising way to overcome these observational challenges is to 
observe at infrared wavelengths.  As has been demonstrated, modern 
infrared spectrographs utilizing molecular gas cells are able to 
obtain RV precisions comparable to very stable optical facilities
\citep[e.g.][]{bean10}.  This wavelength range is advantageous for young 
stars on cool convective tracks since it is much closer to the peak of 
their energy distribution, thereby improving observing efficiency.
More significantly, at these wavelengths the contrast between the 
photosphere and cool star spots is reduced; this
is confirmed by the decline in the photometric variability of spotted
stars with wavelength \citep[e.g.][]{herbst94}.  Since the RV
``noise'' induced by star spots is directly proportional to 
the amplitude of a star's photometric variability \citep{saardonahue97},
the diminished variability at 
longer wavelength observations should help mitigate the noise of 
star spots.  Observations of young spotted stars at both optical and 
infrared wavelenghts appear to confirm this 
prediction \citep{prato08, huelamo08, mahmud11}.

Motivated by the prospects for obtaining high-precision RV 
measurements of young stars at infrared wavelengths, in late 2004 we 
began an ambitious observational program at the W. M. Keck Observatory
(PI: R. White) to search for young planets.  Here we present initial 
results.  In Section 2 we describe the sample selection and in Section 
3 we summarize the acquisition of infrared spectra.  In 
Sections 4 and 5 we describe our method for optimally extracting the 
spectra
and for modeling the spectra to determine precise RVs.  The results of 
this survey, which include the discovery of two spectroscopic binaries 
and one candidate hot Jupiter, are highlighted in Section 6.  In 
Section 7 we determine detection limits for the ensemble sample via
Monte Carlo simulations and quantify the advantage infrared
RVs measurements have over optical measurements for young stars.
Our overall findings are summarized in Section 8.

\section{The Observational Sample: Young Stars and Comparison Field Stars}

Although our entire young star survey includes stars from both star 
forming regions and nearby loose associations, in this first paper 
we focus only on young stars in the $\beta$ Pic and TW Hydrae 
Associations, which have ages of $\sim 12$ Myr and $\sim 8$ Myr, 
respectively \citep{torres08}.  The high quality spectra of these 
nearby, systematically brighter stars were our initial focus as 
we developed techniques for precise RV measurements.  The selection 
criteria for targets within these associations were that the star
have no known companions within 2\farcs0, be somewhat slowly rotating 
($v$sin$i$ $\lesssim$ 15 km/s, if known), show no signs of accretion 
(e.g. TW Hya itself was excluded), and have an M spectral
type.  The spectral type criterion was adopted to ensure that that stars
have strong $^{12}$CO R-branch lines in their spectra, from
which RVs are measured.  Stars were selected from the membership lists 
assembled in \citet{zuckermansong04}; 9 stars from the $\beta$
Pic Association and 9 stars from the TW Hydrae Association met these
criteria.  Wide companions to 2 of these stars have spectral types just 
slightly earlier than M, AG Tri A (K6) and TWA 9A (K5), and were 
therefore also included since their acquisition overhead would be minimal.
Spectroscopic observations of a few of these stars (GJ 871.1 B, TWA 12,
TWA 23) revealed that they were more rapidly rotating than our initial 
selection criterion.  We nevertheless continued to observe these stars 
in order to investigate empirically how the precision degrades with 
projected rotational velocity.

After our observational program began, one of our targets, GJ3305, was 
discovered to be a 0\farcs093 binary \citep{kasper07}, making it a
triple star system \citep{feigelson06}.  It was consequently observed
less often as it was considered a less likely planet host.
We also note that of the 20 young stars in this sample, 14 of 
them (70\%) are members of wide binary star systems, with separations
ranging from 29 to 1967 AU.  These include
AG Tri A and B, GJ 799 A and B, GJ 871.1 A and B, GJ3305, TWA 8A and 8B, 
TWA 9A and 9B, TWA 11B, TWA 13A and 13B.

\begin{deluxetable*}{llc ccc ccc}
\tablecaption{Observed Sample  \label{table1}}
\tablewidth{0pt}
\tablehead{ \colhead{Star}
& \colhead{}
& \colhead{}
& \colhead{$K$}
& \colhead{Mass}
& \colhead{Period} 
& \colhead{}
& \colhead{$v$sin$i$} 
& \colhead{} \\
\colhead{Name}
& \colhead{SpT}
& \colhead{Reference}
& \colhead{($mag$)} 
& \colhead{(M$_\odot$)}
& \colhead{($days$)}
& \colhead{Reference}
& \colhead{($km/s$)} 
& \colhead{Reference}
}
\startdata
\cutinhead{Single Field Stars}
GJ 628  & M3.5	  & H94 & 5.07 & 0.30 &\nodata &\nodata  & $< 1.1$ & D98 \\
GJ 725A & M3 	  & H94 & 4.43 & 0.37 &\nodata &\nodata     & $< 2.8$ & D98 \\
GJ 725A & M3.5	  & H94 & 5.00 & 0.30 &\nodata &\nodata     & $< 2.8$ & D98 \\
\cutinhead{Chromospherically Active Field Stars}
GJ 873 & M3.5     & H94 & 5.30 & 0.30 & 4.38 & M95          &\nodata &\nodata \\
\cutinhead{$\beta$ Pic Stars}
AU Mic & M0       & H96 & 4.53 & 0.73 & 4.87 & M95          & 8.5 & S07 \\
AG Tri A & K6     & S03 & 7.08 & 0.94 & 13.7 & N07          &\nodata &\nodata  \\
AG Tri B & M0     & S03 & 7.92 & 0.73 &\nodata &\nodata     &\nodata &\nodata  \\
GJ 182 & M0       & H96 & 6.26 & 0.73 & 4.56 & M95          &\nodata &\nodata \\
GJ 3305 & M0.5    & H96 & 6.41 & 0.67 & 6.10 & F06     & 5.3 & S07 \\
GJ 799 A & M4.5   & H96 & 5.69 & 0.16 &\nodata &\nodata   & 10.6 & S07 \\
GJ 799 B & M4     & H96 & 5.69 & 0.22 &\nodata &\nodata   & 17.0 & S07 \\
GJ 871.1 A & M4   & S02 & 6.93 & 0.22 & 2.355 & M10     & 14.0 & S07 \\
GJ 871.1 B & M4.5 & S02 & 7.79 & 0.16 &\nodata &\nodata     & 24.3 & S07 \\
HIP 12545 & M0    & S03 & 7.07 & 0.73 & 1.25 & M10     & 9.3 & S07 \\
\cutinhead{TW Hya Stars}
TWA 7 &  M1       & W99 & 6.90 & 0.60 & 5.05 & LC05 &        $< 5$ & S07 \\
TWA 8A & M2       & W99 & 7.43 & 0.51 & 4.65 & LC05 &        $< 5$ & S07 \\
TWA 8B & M5       & W99 & 9.01 & 0.12 & 0.78 & LC05          & 11.2 & S07 \\
TWA 9A & K5       & W99 & 7.85 & 1.01 & 5.10 & LC05          & 11.3 & S07 \\
TWA 9B & M1       & W99 & 9.15 & 0.60 & 3.98 & LC05          & 8.4 & S07 \\
TWA 11B &M2.5     & W99 & 8.35 & 0.44 & \nodata &\nodata     & 12.1 & S07 \\
TWA 12 & M2       & S99 & 8.05 & 0.51 & 3.28 & LC05          & 16.2 & S07 \\
TWA 13A &M1       & S99 & 7.49 & 0.60 & 5.56 & LC05          & 10.5 & S07 \\
TWA 13B &M2       & S99 & 7.46 & 0.51 & 5.35 & LC05          & 14.8 & S07 \\
TWA 23 & M1       & S03 & 7.75 & 0.60 &\nodata &\nodata &\nodata &\nodata \\
\enddata 
\tablecomments{References: 
D98 = Delfosse et al. (1998)
F96 = Feifelson et al. (2006), 
H94 = Henry et al. (1994), 
H96 = Hawley et al. (1996), 
LC05 = Lawson \& Crause (2005), 
M95 = Mathioudakis et al. (9195), 
N07 = Norton et al. (2007),
S02 = Song et al. (2002), 
S03 = Song et al. (2003),
S07 = Sholtz et al. (2007), 
W99 = Webb et al. (1999) }
\end{deluxetable*}

In addition to these 20 young stars, we also included in our sample the 
well studied single M dwarf flare star GJ 873 (EV Lac).  GJ 873 is the 
second brightest M dwarf X-ray source in the \textit{R\"ontgensatellit} 
All-Sky Survey \citep{hunsch99, osten06}, has one of the strongest
surface magnetic field strengths measured on an M dwarf 
\citep{johnskrullvalenti96, reinersbasri07}, and a differential star 
spot covering factor of 4-11\% \citep{abranin98}.  While its activity 
suggests a young age, its dwarf-like gravity and the absence of lithium 
in its atmosphere \citep{pettersen84} indicate it is not pre-main 
sequence.  It's usually classified as a young disk star 
\citep[e.g.][]{osten06} with an approximate age of $10^8$ yr.  The 
motivation for including this star was to assess the
achievable RV precision at infrared wavelengths in the 
limit of extreme surface activity.

For calibration purposes we observed a handful of A spectral type 
stars during each observing run.  The telluric absorption features 
superimposed upon these nearly featureless stellar spectra, at 
infrared wavelengths, are used to characterize the instrumental profile.  
In addition, we observed 3 mid-M dwarf stars (GJ 628, GJ725A and GJ725B)
that have been previously monitored with high precision optical 
RV techniques and shown to have no detectable RV companions \citep{nidever02, 
endl06}; \citet{endl06} measure RV dispersions of 7.4 m/s and 7.1 m/s
for GJ 725A and GJ 725 B, respectively.  These stars
were included to assess the achievable RV precision in the limit
of slowly rotating, inactive stars.

Stellar masses for the young stars are estimated by comparing their 
stellar temperatures to a theoretical isochrone.  To do this we 
assembled stellar spectral types from original references
(see Table \ref{table1}) and assigned temperatures using the
T Tauri-like temperature scale of \citet{luhman03} for stars cooler than 
M0, and the dwarf temperature scale of \citet{kraushillenbrand07} for 
hotter stars.  These temperatures are then compared to the 10 Myr 
isochrone of \citet{baraffe98} with a mixing length of 1.9 above 0.6 
M$_\odot$ and a mixing length of 1.0 at lower masses; this evolutionary
model with the adopted temperature scale is consistent with 
the available dynamical mass constraints for young stars 
\citep{hillenbrandwhite04, mathieu07}.  This yields stellar masses 
that range from 0.12 M$_\odot$ to 1.01 M$_\odot$, with a median mass
of 0.6 M$_\odot$.  Approximate
masses for the field dwarfs are estimated from the absolute
magnitude - mass relations of \citet{henrymccarthy93}, where
the absolute magnitudes have been assigned based on spectral
type relations given in \citet{kraushillenbrand07}.

Table \ref{table1} summarizes the basic properties of the sample 
observed, excluding the A spectral type stars.  The listed properties 
include spectral types, $K$ magnitudes, mass estimates, rotational 
periods, if known, previously reported $v$sin$i$ values, if available, 
and the corresponding references.  The $K$ magnitudes for the field 
stars are taken from \citet{leggett92} whereas the values for 
Association members are from the 2MASS survey, and thus are $K_s$ 
magnitudes. Since the 2MASS survey reports only a combined magnitude for 
GJ799AB, a 2\farcs8 pair, we estimate each component's magnitude
by assuming they are equally bright; the components of this binary 
have the same spectral type.

\section{High Dispersion Infrared Spectra}

Spectroscopic observations of the sample stars were obtained using the 
cross-dispersed infrared echelle spectrograph NIRSPEC \citep{mclean98} 
on the W. M. Keck II telescope.  The observations were obtained
over 12 observing runs in total, spanning from 2004 Nov 16 to 2009 
May 12.  The large number of runs was a consequence of this program 
being completed in concert with a similar program focused on L dwarfs 
\citep{blake07, blake10}.  


All observations were obtained with the 3 pixel (0\farcs432) slit 
in combination with the N7 blocking filter, and approximate
echelle angle 62.65 and grating angle 35.50.  This yielded
7 orders of spectra spanning approximately 1.99 $\mu$m to 2.39
$\mu$m, with gaps between the orders, at a resolving power of
approximately 30,000 (measured as described below). The spectra 
were obtained in pairs at two locations along the slit. Collecting 
observations in this manner provided a nearly simultaneous 
measurement of sky emission and detector bias for both images.

Since the measurement goal of this program was precise RVs,
considerable effort was put into maintaining instrument stability
by not moving any internal components of NIRSPEC while observing.
However, when the NIRSPEC software is first started (at the 
beginning of a night) or when a software crash occurs, 
the motors for the filter wheels and reflection angles must
all be reinitialized, forcing the internal components to move.
Since resetting the reflection angles typically results in a 
slightly different setting, these values were tweaked slightly,
by eye, to return to the primary wavelength setting; this procedure
resulted in the same wavelength coverage to within a few pixels.
Since the spectroscopic properties of this new setting would
be slightly different, we strived to characterize each 
'observational set' by intensively observing one of the A spectral
type stars for each NIRSPEC setting.  We were successful in this
effort for all but one 'observational set'.

\section{Image Reduction and Spectral Extraction}

For each observational set, described above, dark images were 
median combined and this median was subtracted from each
flat-field image.  These subtracted flat-field images were then 
normalized by counts in the central $\sim 10\%$ of the array, and 
then median combined to generate a master flat.  Regions outside the 
illuminated 7 orders were set to unity.  The orders were not 
normalized in the direction of dispersion, a necessary step to 
strictly preserve the counts and the correct relative signal-to-noise 
ratio (SNR); skipping this led to fringing effects that were more 
consistent,
smaller in amplitude, and easier to remove (see below).  All
spectrum images were then divided by the master flat constructed
for its observational set, and then each flattened image had its 
corresponding nod image subtracted from it to remove the sky 
emission, detector bias and dark current. This yielded 2 reduced 
spectrum images for each epoch.

Prior to spectral extraction, the 7 spectral orders were located 
on the reduced spectrum image automatically by finding maxima in 
smoothed spatial profiles of each order (i.e. perpendicular 
to the dispersion direction).  Gaussian functions were then fit to 
these spatial profiles in binned segments of 10-columns along the 
orders.  The 
orders were then traced by fitting a low order polynomial to the 
peak position of these Gaussians, with 3$\sigma$ outliers excluded
from the fit.  Next, regions that were more than 3 sigma above 
or below the best fit polynomial defining the center of each order 
were masked and excluded
from any further analysis.  Masking the orders in this way 
objectively accounted for variations in the spatial width of the
spectra, usually caused by variations in the seeing, and proved 
vital for successful extraction of pair subtracted spectra only 
10\farcs0 apart.

Spectra were extracted from the masked images using a modified
version of the 'optimal extraction' procedure described in 
\citet{horne86}.  Unlike standard extraction techniques which 
simply sum pixel values over a specified range, optimal 
extraction sums pixels weighted by the variance of an assumed
smoothly varying spatial profile that can be defined either 
perpendicular to or parallel to the dispersion direction.  
This minimizes the noisy contributions of profile wings and 
allows significant deviates from the profile (e.g. hot 
pixels and cosmic rays) to be easily identified and removed
(see Piskunov \& Valenti 2002 for this and other issues 
related to the optimal extraction of cross-dispersed echelle 
spectra).  After considerable experimentation, we adopted 
spatial profiles defined parallel to the dispersion direction 
and used a Gaussian plus second order polynomial to model these.
The variance of the profile was determined from each pair 
subtracted flat-fielded spectrum image.  We note that this is 
slightly different from the recommendation of \citet{horne86} 
to use a bias subtracted flat-fielded spectrum image; as with 
most infrared observations, the bias is only removed as a part 
of pair subtraction.  The addition of these bias counts produced 
slightly larger variance estimates, but this yielded no 
noticeable effect on our relatively high SNR spectra.  Future 
studies
may consider obtaining bias images throughout the night in an
attempt to determine a more accurate variance image.


An example of an optimally extracted spectrum is shown in Figure 
\ref{fig1}.  One feature not accounted for in these extractions 
is interference fringing.  Although some investigators have 
developed prescriptions to remove this via Fourier filtering
\citep{deming05}, we accounted for this feature in the spectral 
modeling process.

\section{Modeling the Spectra and Precise Radial Velocity Measurements}

\subsection{Assumptions}

Precise RVs were extracted from the observations by constructing
a detailed model of each spectrum.  Although a wide variety of 
modeling prescriptions were investigated with our data to optimize
this, the prescription described here is the one that yielded the 
most consistent RVs for 15+ epochs of 3 slowly rotating M dwarfs 
with no known planets (GJ 628, GJ725A and GJ725B).


The modeled spectra were constructed from an ultra-high resolution
KPNO/FTS telluric spectrum, extracted from observations of the Sun 
\citep{livingstonwallace91}, and synthetically generated stellar 
spectra, computed from updated and improved NextGen models 
\citep{hauschildt99}, constructed and provided by coauthor
T. Barman.  Although the $^{12}CO$ bandhead and R branch 
lines are temperature and surface gravity dependent, we did not
determine these parameters from our spectra; attempts to do this
yielded results that were only marginally consistent with previous
values considered to be more accurate (e.g. from optical spectra).  
Instead, we used a synthetically generated spectrum 
that most closely matched each star's spectral type and expected
surface gravity.  All field stars (GJ 628, GJ725A, GJ725B, GJ 873) 
were assumed to have a log(g) = 4.8 dex, while the young stars were
assumed to have a log(g) = 4.2 dex, consistent with recent log(g)
measurements \citep{mentuch08}.  The stellar properties of the 
spectral templates used for each star are listed in Table 
\ref{table2}.

Each observation was modeled by combining a telluric and a 
synthetic spectrum, parametrized by 10 free parameters.  Four 
parameters determine instrumental properties - the wavelength 
solution was parametrized by a quadratic polynomial, 
characteristic of slightly curved spectral orders, and the 
instrumental spectral profile was modeled as a single best-fit 
Gaussian for the entire order.  The 
remaining six parameters 
characterize the spectral properties, including the depth of 
telluric features, the depth of stellar features, the star's 
projected rotational velocity ($v$sin$i$), a linear (2 parameter) 
continuum normalization offset, and finally, the star's RV.  The 
limb darkening coefficient used to calculate rotationally broadened 
profiles, which was \textit{not} considered a free parameter in these 
calculations, was held fixed at 0.6 in all cases.

\subsection{Implementation}

\begin{figure}
\epsscale{1.15}
\plotone{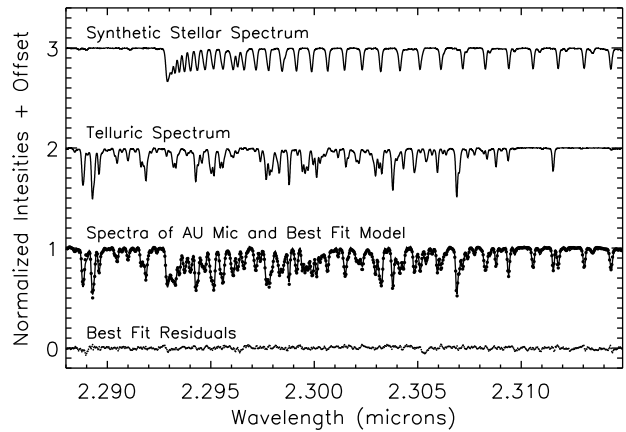}
\caption{Illustration of the procedure to derive precise radial velocities 
using NIRSPEC data.  A telluric spectrum (\textit{top spectrum}) is combined
with a synthetic stellar spectrum (\textit{2nd spectrum}) to construct a
model spectrum (\textit{3rd spectrum, dotted line}; difficult to see) that
is optimized to match an observed spectrum of AU Mic (\textit{3rd spectrum, 
solid line}).  The normalized dispersion of the best fit residuals 
(\textit{4th spectrum}) is 1.5\%, and is characteristic of the majority of 
our fits. \label{fig1}}
\end{figure}

Only 1 of the 7 orders acquired (NIRSPEC order \#33) was modeled; this
is one of the few wavelength regions which has a sufficiently rich 
array of both the stellar and telluric absorption features, permitting
precise RV calibration.  The usable region of the observed order spans 
approximately 270 \AA, from 2.288 $\mu$m to 3.315 $\mu$m (Figure \ref{fig1}). 

The spectral modeling proceeded in essentially 3 stages.  First, the 
A-type star spectra were modeled to construct initial estimates of the 
wavelength solution and instrumental profile for each observational 
set.  Since the spectra of A type stars are essentially featureless
over this wavelength range, their spectra were modeled as simple
telluric spectra, and 3 of the 10 parameters of the general fit were
omitted (stellar absorption depth, $v$sin$i$, and RV).  The best fit 
was determined by subtracting the model spectrum from the observed 
spectrum and applying a Hamming filter to this difference to remove 
the residual fringing pattern.  Optimization was achieved by an iterative
process that minimized the variance-weighted reduced $\chi2$ of 
these filtered residuals, using the prescription described below.
Parameters determined for multiple A star observations within a 
set were averaged, weighted by their $\chi2$ values, to determine the 
best "first guess" values for that set.

The average best fit instrumental profile determined from the best fit
models of the A star spectra
corresponded to a resolving power of $\sim$ 30,000, with a dispersion 
of $\sim 2,000$.  This is larger than expected for this instrumental 
setup, which is predicted to yield resolving powers of $\sim 25,000$ 
\citep{mclean98}.  Although we are unsure of the discrepancy, it may 
be attributable to the excellent seeing at Mauna Kea and starlight 
not fully filling the 0\farcs432 slit.

Each late-type stellar spectrum was modeled by adopting and 
holding fixed its set's estimates of the linear and quadratic terms
of the wavelength solution.  All other parameters were determined 
via an iterative fitting process, but at all times the parameters were
nevertheless restricted to realistic ranges. In the first iteration the 
0th order component of the wavelength solution was optimized by allowing 
it to vary; the parameter describing the instrument profile was adopted 
from its set's estimate. The second iteration optimized the instrument 
profile by allowing it to vary.  The third iteration was used to 
determine the stellar $v$sin$i$ by allowing it to vary.  Once complete, 
a chi-square weighted mean of the best-fit $v$sin$i$ value was computed 
and adopted for the star. In the final iteration only the 5 remaining 
parameters (2 normalization parameters, telluric depths, stellar depths, 
RV) were allowed to vary.

For the 1 observational set without any A star observations (Section 3),
the target stars of that set were modeled using the average A star
parameters from all other sets as a first guess.  The analysis then
proceeded as described above.  The agreement of measured stellar properties
(e.g. RVs, $v$sin$i$) from this set and other sets suggested the lack
of A stars did not compromise the analysis in this case, likely because 
of the nearly identical observing setups carefully determined each time.

The specific prescription used to determine multiple-parameter best
fits can have a significant impact on the results.  
In this first study we adopted a minimization of the filtered difference
spectra based on the downhill simplex method of Nelder \& Mead (1965),
as implemented by Press et al. (1992)\footnote{The version of this
multiple-parameter optimization available in the IDL Astronomy User's
Library, called \textit{AMEOBA}, 
must be converted to double-precision to converge self-consistently.}.
Our implementation of this prescription enforced user-specified 
limits on each parameter by restarting an iteration with new 
parameters if any were out of bounds.  The minimization proceeded 
until both the best fit $\chi2$ dropped by less than one percent 
and all parameters remain consistent with recent values.

Figure \ref{fig1} shows an example model spectrum for comparison with
the observed spectrum of the young star AU Mic.  The residuals to the fit 
have a normalized dispersion of 1.5\%, which is typical for the majority of 
modeled
spectra.  We note that the results from this last stage yield absolute RVs, 
since they are determined relative to \textit{at rest} telluric absorption 
features.  These values are then converted to barycentric velocities 
using a correction prescription accurate to $\sim 1$ 
m/s (G. Basri; priv. comm.).  For the interested reader, we note that 
\citet{seifahrtkaufl08}, \citet{blake10}, \citet{bean10}, 
\citet{crockett11} and \citet{muirhead11} provide detailed 
descriptions of complementary methods to determine precise RVs with 
nod-subtracted infrared spectra.

%


\section{Results}

In Table 2 we summarize the results of our spectroscopic modeling effort
to extract RVs.  The second column lists the synthetic model parameters 
(T$_{eff}$, log(g) ) assigned in the spectral fits to each star, and the 
third column lists the determined $v$sin$i$ values.  Uncertainties in the
$v$sin$i$ values are set to be the standard deviation of the $v$sin$i$
estimates using all epochs for each star and thus are not necessarily 
free of systematic effect. The very large uncertainty for TWA 23 may be a 
consequence of its spectroscopic companion (Section 6.2)
broadening its line profiles in some epochs.

The fourth column in Table 2
lists the mean RV and the uncertainty in this mean ($\sigma / \sqrt{N}$), 
calculated from all observations.  The uncertainties in the mean range 
from 12 m/s to 65 m/s for single stars; these are some of the most
precise absolute system RV measurements ever for young stars.
The fifth and sixth columns 
list the number of observations ($N$) and the standard deviation 
($\sigma$) of the RV measurements extracted from these observations.
In the following analysis, we include GJ 873 in the statistics 
determined for young stars.

\begin{figure}
\epsscale{1.2}
\plotone{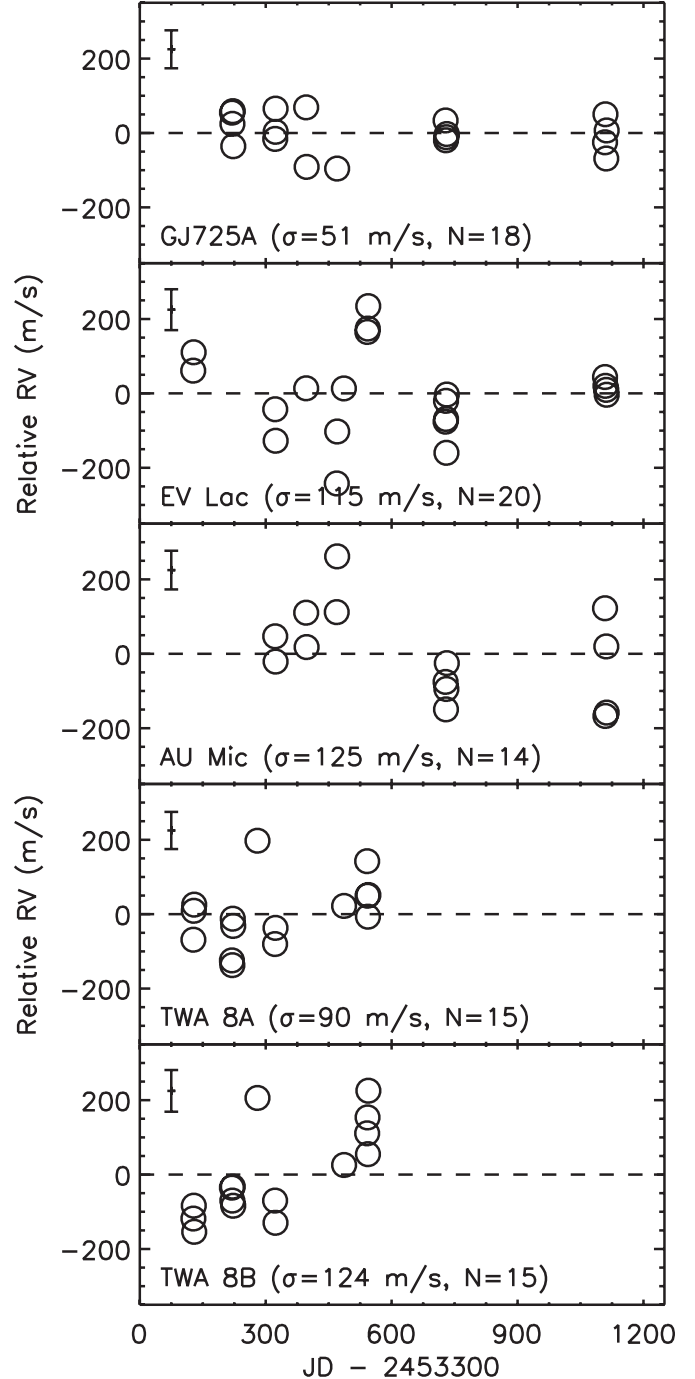}
\caption{RV curves for the slowly rotating single field star GJ725 A, 
the chromospherically active star GJ 873, the debris disk host AU Mic,
and the components of the wide binary TWA 8; average errorbars are
illustrated in the upper left of each panel.  The dispersions of the
field stars, like GJ 725 A, demonstrate an observing precision of 
$\sim 50$ m/s using high SNR spectra of slowly rotating
inactive stars.  The larger dispersions of other stars are primarily a
consequence of enhanced stellar activity.
\label{rv_singles}}
\end{figure}

In the Appendix, RV measurements from each epoch for all stars are 
provided.  The calculation of the measurement uncertainties 
are describe below.  Figure \ref{rv_singles} illustrates several 
RV curves, including the field star GJ725 A (M3.5), 
the flare star GJ 873 (M3.5), the debris disk host AU Mic, and 
the components of the wide binary TWA 8.

\begin{deluxetable*}{llllrrccll}
\tablecaption{Survey Results \label{table2}}
\tablewidth{0pt}
\tablehead{ \colhead{Star}
& \colhead{Model}
& \colhead{$v$sin$i$} 
& \colhead{$<RV>$}
& \colhead{}
& \colhead{$\sigma_{obs}$} 
& \colhead{3d}
& \colhead{10d}
& \colhead{30d}
& \colhead{100d} \\
\colhead{Name}
& \colhead{T$_{eff}$,\,log(g)}
& \colhead{($km/s$)} 
& \colhead{($m/s$)} 
& \colhead{N}
& \colhead{($m/s$)}
& \colhead{($M_{J}$)}
& \colhead{($M_{J}$)}
& \colhead{($M_{J}$)}
& \colhead{($M_{J}$)} \\
}
\startdata
\cutinhead{Single Field Stars}
GJ 628    & 3400, 4.8  & 1.7$\pm$0.1  & $-21113 \pm 15$ & 13 &  55 & 3.7 & 5.3 & 8.7 & 10.5 \\
GJ 725A   & 3400, 4.8  & 1.8$\pm$0.1  & $-611\pm 12$    & 18 &  51 & 5.0 & 6.9 &  13 & 16 \\
GJ 725B   & 3400, 4.8  & 1.8$\pm$0.1  & $+1321 \pm 13$  & 18 &  53 & 3.7 & 5.9 & 9.2 & 13 \\
\cutinhead{Chromospherically Active Field Stars}
GJ 873    & 3400, 4.8  & 4.7$\pm$0.1  & $+545 \pm 5$    & 20 & 115 & 5.1 & 7.9 &  12 & 16 \\
\cutinhead{$\beta$ Pic Stars}
AU Mic    & 3800, 4.2  & 8.7$\pm$0.2  & $-4130 \pm 33$  & 14 & 125 &  11 &  17 &  23 & 32 \\
AG Tri A  & 3800, 4.2  & 4.7$\pm$0.1  & $+6743 \pm 26$  & 14 &  98 &  10 &  17 &  22 & 33 \\
AG Tri B  & 3800, 4.2  & 5.0$\pm$0.1  & $+5961 \pm 35$  & 14 & 132 & 9.9 &  14 &  20 & 29 \\
GJ 182    & 3800, 4.2  & 9.4$\pm$0.7  & $+19818 \pm 42$ &  6 & 103 &  12 &  19 &  25 & 54 \\
GJ 3305   & 3800, 4.2  & 5.7$\pm$0.1  & [$+20625$]      &  5 & 457 & \nodata & \nodata & \nodata & \nodata \\
GJ 799 A  & 3800, 4.2  & 9.6$\pm$0.6  & $-3727 \pm 40$  & 14 & 151 & 3.9 & 5.9 &  8.2 & 12 \\ 
GJ 799 B  & 3200, 4.2  & 14.7$\pm$0.4 & $-5126 \pm 50$  & 13 & 179 & 6.0 & 7.7 &  12 & 18 \\ 
GJ 871.1 A& 3200, 4.2  & 13.9$\pm$0.5 & $+3087 \pm 36$  & 14 & 134 & 5.7 & 8.6 &  12 & 19 \\ 
GJ 871.1 B& 3200, 4.2  & 22.7$\pm$0.3 & $+2031 \pm 44$  & 15 & 169 & 5.2 & 7.3 &  10 & 15 \\ 
HIP 12545 & 3800, 4.2  & 8.7$\pm$0.2  & $+8253 \pm 48$  & 14 & 179 &  12 &  17 &  23 & 33 \\ 
\cutinhead{TW Hya Stars}
TWA 7     & 3800, 4.2  & 4.7$\pm$0.1  & $+12446 \pm 14$ & 12 &  94 & 7.4 &  12 &  19 & 25 \\ 
TWA 8A    & 3600, 4.2  & 4.7$\pm$0.1  & $+8679 \pm 23$  & 15 &  90 & 6.7 &  10 &  15 & 22 \\ 
TWA 8B    & 3400, 4.2  & 10.5$\pm$0.3 & $+8607 \pm 32$  & 15 & 123 & 3.6 & 4.5 & 7.1 & 11 \\ 
TWA 9A    & 4000, 4.2  & 10.4$\pm$0.3 & $+11649 \pm 21$ & 11 &  71 &  15 &  25 &  30 & 47 \\ 
TWA 9B    & 3800, 4.2  & 8.6$\pm$0.3  & $+12279 \pm 27$ & 11 &  90 & 8.8 &  15 &  20 & 29 \\ 
TWA 11B   & 3600, 4.2  & 12.0$\pm$0.2 & $+8923 \pm 59$  & 11 & 191 &  10 &  12 &  19 & 29 \\ 
TWA 12    & 3600, 4.2  & 17.0$\pm$0.3 & $+12498 \pm 52$ & 12 & 181 &  11 &  14 &  20 & 31 \\ 
TWA 13A   & 3800, 4.2  & 11.1$\pm$0.3 & $+11668 \pm 65$ & 14 & 242 & \nodata & \nodata & \nodata & \nodata \\ 
TWA 13B   & 3600, 4.2  & 10.7$\pm$0.3 & $+12075 \pm 43$ & 13 & 149 & 9.0 &  13 &  18 & 26 \\ 
TWA 23    & 3800, 4.2  & 20.5$\pm$5.5 & [$+6520$]       & 15 &2425 & \nodata & \nodata & \nodata & \nodata \\ 
\enddata
\end{deluxetable*}

In the following sections the ensemble RV dispersions, as measured by
the standard deviations, are used to assess empirically the precision 
with which RVs can be measured and to identify RV variables.

\subsection{Error Analysis}

For single stars, the RV dispersions are assumed to stem from 3 causes 
of variability: 
(1) a theoretical photon noise error ($\sigma_{phot}$) based on the 
SNR
of the observed spectrum and the number and shape of the features in 
best-fit synthetic and telluric spectra, 
(2) an instrumental error ($\sigma_{inst}$) based on the 
characteristics of NIRSPEC spectra, and (3) an intrinsic 
stellar error ($\sigma_{stel}$) caused by stellar activity.
Although the second and third causes are likely attributable to 
non-stochastic events (e.g. variable bias, fringing or stellar flares, 
star spots, respectively), we nevertheless assume that these variability
sources add in quadrature to produce the observed dispersions
($\sigma_{obs}2 = \sigma_{phot}2 + \sigma_{inst}2 + \sigma_{stel}2$).

Observations of the 3 slowly rotating single field stars are used to 
quantify the 
magnitude of $\sigma_{inst}$.  As noted in Section 2, previous precise 
RV measurements of these stars show that their $\sigma_{stel}$ is 
small ($< 10$ m/s); we assign values of zero for all 3 stars.
The $\sigma_{phot}$ values are calculated for each pair 
averaged observation using the SNR and a prescription similar to 
that outlined in \citet{butler96}. Essentially this error term is
computed as the photon-limited Doppler error of the best fit synthetic
spectrum added in quadrature with the photon-limited wavelength error 
of the best fit telluric spectrum.  Because we successfully
achieved similar high SNRs for all observations, the median 
pair-averaged $\sigma_{phot}$ for the 3 field stars are
similar (26.2 m/s, 26.4 m/s, 27.1 m/s).  We use these to 
estimate $\sigma_{inst}$ by removing the photon noise error 
$\sigma_{phot}$ from the observed dispersion values 
($\sigma_{inst}2 = \sigma_{obs}2 - \sigma_{phot}2$).  The 
$\sigma_{inst}$ values for these 3 stars are consistent to 
within 4 m/s, with a median of 46 m/s.  These values are only 
slightly smaller than their observed dispersions, as expected
for high SNR spectra of inactive stars.  We note that a possible
error term not accounted for in these calculations is the 
stability of the telluric features used for calibration.  However,
optical \citep{cochran88} and infrared \citep{seifahrt10} RV 
studies conducted with other facilities have shown these features 
to be stable to $\lesssim 10$ m/s; their stability is thus not a 
limiting factor for precisely measured RVs with NIRSPEC.
These results imply that
achieving RV precisions significantly better than $\sim 40$ m/s 
with NIRSPEC will be challenging unless some methods are developed 
to account for the instrumental noise, which we presume is 
attributable to the features of its 1990s generation array 
\citep[][e.g. read noise, fringing]{mclean98}.  It is also
possible that some improvements may be realized using 
alternative spectral modeling techniques that incorporate 
template spectra instead of synthetic spectra and/or multiple-Gaussian
instrumental profiles.  Errors inherent to our modeling prescription
are incorporated into the instrument error reported here. 
Regardless, 
this precision is considerably better than intended in the 
design specifications, and a credit to the talent of the 
instrument team.

Measurement errors are assigned to each observation (listed in the 
Appendix) by calculating $\sigma_{phot}$ for each pair-averaged 
observation, and adding that in quadrature with an assumed constant 
$\sigma_{inst}$ of 46 m/s.  These errors do not account for any 
possible errors due to stellar activity ($\sigma_{stel}$).
The RV errors listed 
in the Appendix for the 21 young stars range from 46 
m/s to 79 m/s with a median values of 53 m/s.  The $\sigma_{obs}$ 
values are nevertheless significantly larger than these error
estimates, ranging from 71 - 197 m/s for the single stars 
(identified below).  This implies a significant contribution from
the $\sigma_{stel}$ to the observed dispersions.

Since stellar activity is known to be correlated with rotational 
velocity\citep[e.g.][]{baliunas95}, we investigate the relation 
between the 
observed dispersions and projected rotational velocities in Figure 
\ref{vsini_sigma_quadratic}.  Although there is considerable 
spread in the dispersion values near a $v$sin$i = 10$ km/s, the 2
quantities nevertheless appear to be correlated.  Stars with 
$v$sin$i$ values less than 12 km/s have a median dispersion of 
115 m/s while those rotating more rapidly have a median
dispersion of 179 m/s.


\subsection{New and Candidate RV Variables}

\begin{figure}
\epsscale{1.2}
\plotone{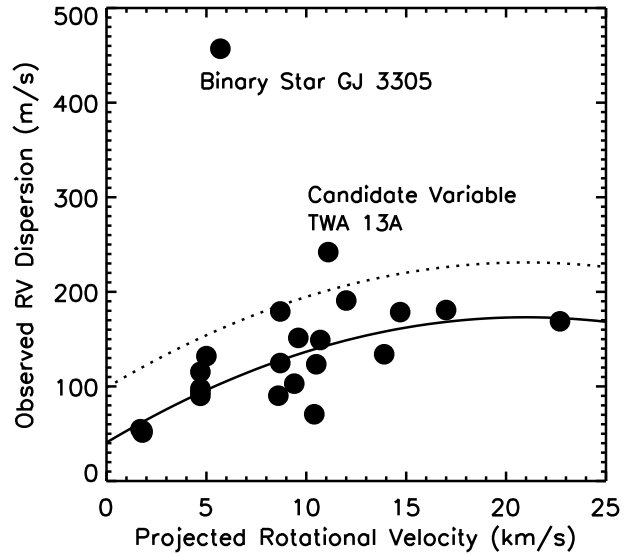}
\caption{RV dispersion versus projected rotational velocities;
with a dispersion of 2425 m/s, TWA 23 is not illustrated on this plot.
A quadratic function is fit via least squares to the data (\textit{solid line})
and systems with dispersions 2$\sigma$ above this best fit (\textit{dotted line})
are considered candidate RV variables.
\label{vsini_sigma_quadratic}}
\end{figure}

We use the apparent relation between the observed RV dispersions 
($\sigma_{obs}$) and $v$sin$i$ values illustrated in Figure 
\ref{vsini_sigma_quadratic} to objectively identify possible
RV variables.  To do this, a quadratic polynomial is fit to the data 
iteratively; stars with dispersions 2$\sigma$ above the best fit are
identified as possible RV variables and subsequently excluded from the 
fit.  This identified 3 stars as possible variable stars, TWA 23, 
GJ 3305, and TWA 13A.  We note that these same stars would be 
identified using a linear fit to the data, but we adopt and 
illustrate in Figure \ref{vsini_sigma_quadratic} a quadratic fit
which can account for saturation in the dispersion,
if present, given the limited range of dispersions observed (see below).

\begin{figure}
\epsscale{1.2}
\plotone{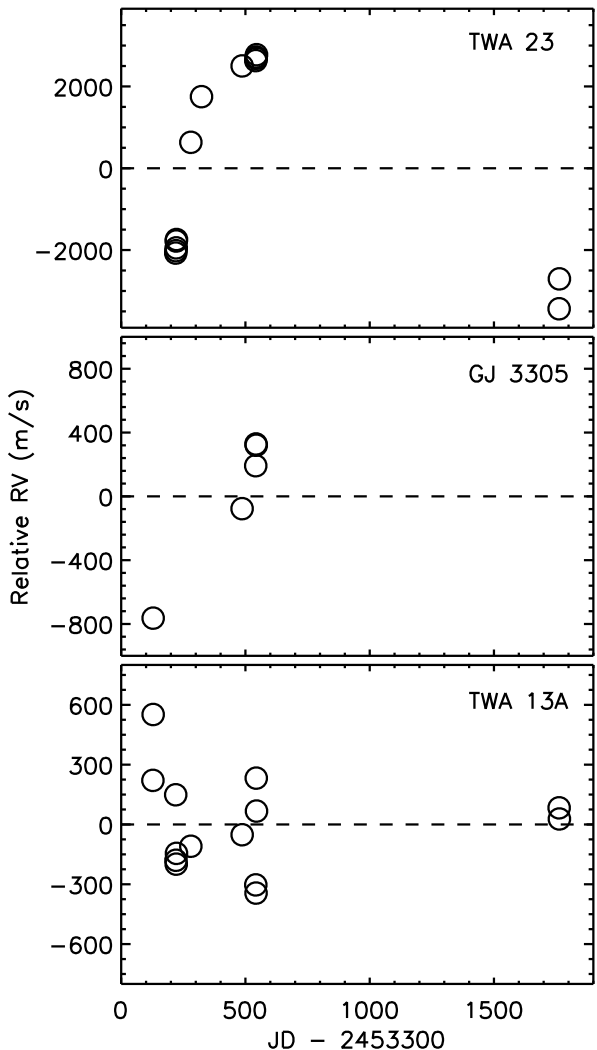}
\caption{RV curves for the new binary stars TWA 23 and GJ 
3305, and the candidate RV variable TWA 13A; typical uncertainties in
these measurements are 50-60 m/s.  The long
time-scale ($> 1$ yr) RV trends of TWA 23 and GJ 3305 are indicative of 
low mass stellar, or possibly brown dwarf companions.  In the case of 
GJ 3305, the motion is likely due to its spatially resolved companion
\citep{kasper07}.  The large amplitude, short-timescale variations of 
TWA 13A are indicative of a short-period gas giant planet (e.g. a hot 
Jupiter), but the data are insufficient to confirm this.
\label{rv_variables}}
\end{figure}

\begin{figure}
\epsscale{1.0}
\plotone{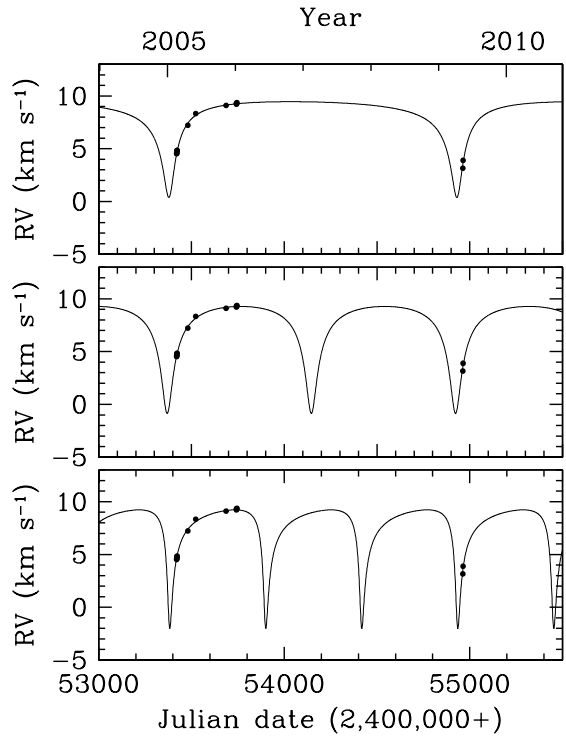}
\caption{Three plausible orbits for the binary star TWA 23, with periods
of 1552 d, 777 d, and 517 d.
\label{twa23_orbits}}
\end{figure}

Two stars have dispersions well above their predicted dispersion for
their $v$sin$i$, TWA 23 ($\sigma_{obs}$ = 2425 m/s) and GJ 3305 ($\sigma_{obs} 
=$ 457 m/s; Table 2).  These large dispersions,
in combination with their smoothly accelerating RV variations illustrated
in Figure \ref{rv_variables} confidently indicate that these stars are 
spectroscopic 
binaries.  In the case of GJ 3305, the observed variations are likely due 
to its 0\farcs093 companion \citep{kasper07}.  The discovery of this 
spatially resolved companion mid-way through our observational program is 
why we have fewer observations of it; it was considered a less likely place 
to find a young planet.  
In the case of TWA 23, the follow-up observations in 2009 helped to 
establish an upper limit on its orbital period, but its orbit is still
underdetermined.  In Figure \ref{twa23_orbits} we show 3 plausible orbital 
solutions for TWA 23, determined using a standard iterative non-linear 
least-squares method (Levenberg-Marquardt, outlined in Press et al. 1992, 
p. 650); with rms residuals of $\sim 0.2$ km/s in all cases, the 3 fits 
are equally good.  The orbital periods of these 3 fits are 517 d, 777 d, 
and 1552 d with amplitudes of 4 km/s, 5 km/s, and 5 km/s.  As with GJ 3305,
its companion is more likely to have a stellar mass.  Nevertheless, the 
discovery of these relatively low amplitude spectroscopic binaries 
highlights the potential for new discoveries that high-precision infrared 
RVs have when applied to young stars.

The RV dispersion of one other star stands out as being marginally 
above (3.4$\sigma$) the best-fit relation, TWA 13A.  Inspection of its RV
curve shown in Figure \ref{rv_variables} reveals large amplitude variations 
($>200$ m/s) on relatively short (1-2 day) timescales.  These types of 
variations are the type expected if TWA 13A is orbited by a hot Jupiter-like 
companion.  Unfortunately our sampling and precision are insufficient to 
confirm this possibility.  We classify it as a candidate 
variable\footnote{Prior to the follow-up observations in 2009, TWA 13A was 
even further above the best fit relation ($\sigma_{obs} =$ 261 m/s versus 
242 m/s); the new observations 
near its mean reduced its overall dispersion.  However, since it was 
initially identified as a candidate variable for follow-up observations, 
and remains somewhat discrepant with its expected precision, we continue
to classify it as a candidate RV variable.}

It is possible that the larger-than-expected RV dispersion of TWA 13A is
caused by either instrumental errors or stellar activity being larger 
than expected for this star during these observations; these sources of 
variation are harder to predict and account for.  As a possible test for 
the former, we compared the RV dispersion of TWA 13A to that of its 
companion TWA 13B.  Both stars are equally bright ($K = 7.5$) and in 13 
of TWA 13A's 14 observations, both stars were observed either immediately 
before or after one another.  If the NIRSPEC detector was especially 
noisy during these observations, it likely would have affected observations
of TWA 13B as well.  However, the observed dispersion of TWA 13B 
($\sigma_{obs} = 149$ m/s) agrees almost exactly with that predicted from 
its $v$sin$i$ (142 m/s).  This strengthens the case that TWA 13A is a RV 
variable star, but unfortunately it is not yet possible to determine if 
the variations are caused by a companion or stellar activity.

\section{Discussion}

\subsection{Companion Detection Limits}

Companion mass detection limits are determined via Monte Carlo simulations.
Since this survey is most sensitive to short period companions, we set limits 
only at representative orbital periods, namely 3, 10, 30 and 100 days.  The
shortest period is choosen to be near the peak in the orbital period distribution
of hot Jupiters while the other periods are chosen to demonstrate how the 
sensitivity declines with increasing orbital period, given the precision and 
temporal sampling of this survey.  For each star, a set of 10,000 orbits is 
generated with random inclinations, phases, and companions spanning a broad 
range of planetary and brown dwarf masses.  For discussion purposes, we refer to
companions with masses $\le 13$ M$_{Jupiter}$ as planets and more massive 
companions (up to $\sim 65$ M$_{Jupiter}$) as brown dwarfs.  All orbits are 
assumed to be circular,
as would be expected if these companions migrated to these locations via disk
interactions \citep[e.g.][]{lin96}.  In combination with the stellar masses 
listed in Table \ref{table1}, these orbits are used to predict a set of RV 
measurements with the same temporal sampling as each star's set of observations.
An assumed Gaussian error of 53 m/s, equal to the median observational error
above ($\sigma_{phot}2 + \sigma_{inst}2$), is added to these measurements; we
note however that adding these relatively small errors has no significant 
effect upon the results.
Conservative detection limits are then set by determining the companion 
mass that induces a RV dispersion that is 2$\sigma$ greater than the 
expected dispersion for the star's $v$sin$i$ (see Figure 
\ref{vsini_sigma_quadratic}) 99\% of the time; this is consistent with
the criteria used to identify candidate variables such as TWA 13A.

The companion detection limits determined from this prescription at 3, 10,
30, and 100 days are listed for all single stars in Table \ref{table2}.  
The 3 field stars, which have the smallest RV dispersions, have the lowest 
companion mass detection limits.  On average, these are
4.1, 6.0, 10, and 13 M$_{Jupiter}$ for the above 4 orbital periods, 
respectively.  These observations confirm that
these stars have no short-period massive planet or brown dwarf companions
at any orbial orientation except possibly nearly face-on.

In assessing the young stars detection limits, we exclude the known or 
candidate RV variables and use only results for stars with 11 or more 
observations (thus excluding GJ 182 with only 6 observations).
For this 18 star subsample, the average detection limits are 8.5, 13, 17, and 
26 M$_{Jupiter}$ at the above 4 orbital periods, respectively.  
The lowest mass stars within this subset have the lowest companion
detection limits, as expected since the detection limit scales as 
$M_{star}^{2/3}$.  For example, the 0.12 M$_\odot$ star TWA 8B has the 
smallest detection limits of 3.6, 4.5, 7.1, and 10.6 M$_{Jupiter}$, 
respectively, despite having an observed dispersion of 123 m/s.
Overall, the observations only exclude the presence of the very 
massive hot ($P < 10$ d) Jupiters.  However, the observations 
confidently exlude all hot brown dwarf companions and all but the 
lowest mass 'warm' ($10 < P < 100$ d) brown dwarf companions.

The RV measurements would translate to much stricter companion detection 
limits if the companion's orbital plane is assumed to be edge-on.
This assumption is plausible
for the young star AU Mic that is known to harbor an edge-on circumstellar
disk \citep[e.g.][]{kalas04}.  At an assumed age of 12 Myr, its M0 spectral 
type corresponds to a stellar mass of 0.73 M$_\odot$.  For this star, 
we calculate companion detection limits using the same Monte Carlo 
simulations described above, but assuming an edge-on orientation for 
the orbit.  This yields companion detection limits of 1.8, 2.5, 3.9
and 5.2 M$_{Jupiter}$ at the above 4 periods, respectively.  These 
detection limits are consistent with the lack of any transiting 
Jupiter-sized planets found in this system \citep{hebb07}.  If massive 
Jupiters do exist within the AU Mic system, they have not migrated inwards 
in its $\sim 12$ Myr lifetime.

\subsection{The Effect of Stellar Activity on Precision RV Measurements}

\begin{figure}
\epsscale{1.2}
\plotone{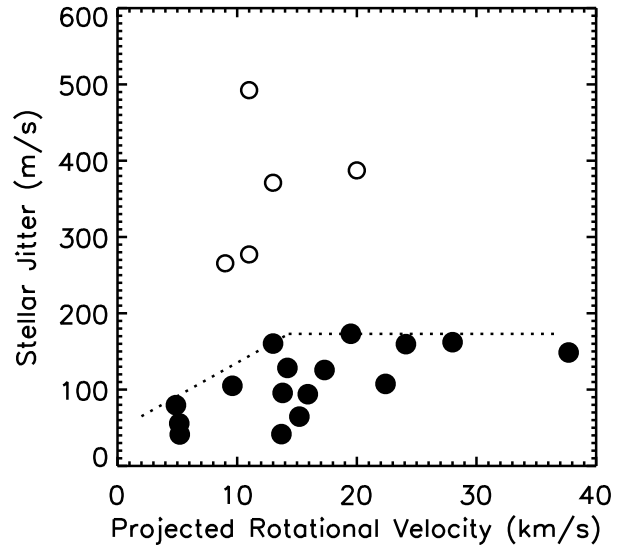}
\caption{
Stellar jitter versus projected rotational velocity.  The 
stellar jitter at infrared wavelengths (\textit{solid circles}) 
is significantly less than at optical wavelengths (\textit{empty
circles}), as expected if it is primarily caused by cool star 
spots.  The \textit{dotted line} illustrates the upper boundary 
of infrared stellar jitter values, and illustrates the tentative
evidence for saturation at $\sim 180$ m/s for $v$sin$i$ values
above 12 km/s.
\label{stellar_jitter}}
\end{figure}

One of the primary motivations for conducting this RV survey of young
stars at infrared wavelengths, as opposed to optical wavelengths, was 
to mitigate the RV variations induced by stellar activity; we subsequently
refer to this as stellar jitter.  To investigate the success of this, we 
compare estimates of the stellar jitter of our infrared observations to 
the estimates from optical observations.  The infrared stellar jitter
is estimated by subtracting, in quadrature, the instrumental and average 
theoretical errors from the observed dispersion for each star 
($\sigma_{stel}^2 = \sigma_{obs}^2 - \sigma_{inst}^2 - \sigma_{phot}^2$).  
This yielded values ranging from 36 m/s to 181 m/s, with a median value of 
113 m/s; these are illustrated in Figure \ref{stellar_jitter} versus 
$v$sin$i$.

The values are then compared to the stellar jitter values calculated
from RV measurements in \citet{paulsonyelda06}, who
obtained precise optical RV measurements of 6 members of the $\beta$ Pic 
Association, including 2 stars in the present survey here, AU Mic and 
GJ 3305.  GJ 3305 is not included in these comparisons, however, because 
we identify it as a spectroscopic binary star.  \citet{paulsonyelda06}
obtained precise RVs using the telluric oxygen band at 6900 \AA\, with
an instrumental precision of $40$ m/s.  As with the infrared
measurements, the stellar jitter is estimated by subtracting, in 
quadrature, the theoretical and instrumental uncertainties from the 
observed velocity dispersions.  These results are plotted in Figure 
\ref{stellar_jitter} versus $v$sin$i$, and range from 270 m/s to 490 
m/s, with a mean of of 360 m/s.  A comparison of these suggests that
the stellar jitter at infrared ($K$-band) is approximately 3 
times less than at optical wavelengths ($R$-band).  This is consistent with
the reductions meausured by \citet{huelamo08} for the similar age star 
TW Hya.  The recent study by \citet{mahmud11} reports similar reductions 
between infrared and optical jitter for the younger T Tauri stars DN Tau 
and Hubble I 4, and slightly larger reductions (a factor 4-5) for V827 
Tau and V836 Tau.

Although the binary star GJ 3305 is not included in these comparisons, 
we note that \citet{paulsonyelda06} do not identify it as a spectroscopic
binary, despite having observed it 11 times on 4 different nights.
We attribute this to their limited sampling, its apparently low
amplitude RV variations, and the larger stellar jitter associated with
optical observations of young stars.  We also note that we can not
confirm the claim of \citet{song03} that HIP 12545 is a single-lined 
spectroscopic binary.  Although they do not report individual measurements, 
they claim to have measured large 
(20 km/s) RV variations for this star using on high dispersion optical
spectra.  Our 14 measurements of this star spanning from 2004 November 
19 to 2006 July 6 have a dispersion of only 0.18 km/s.

\begin{figure}
\epsscale{1.2}
\plotone{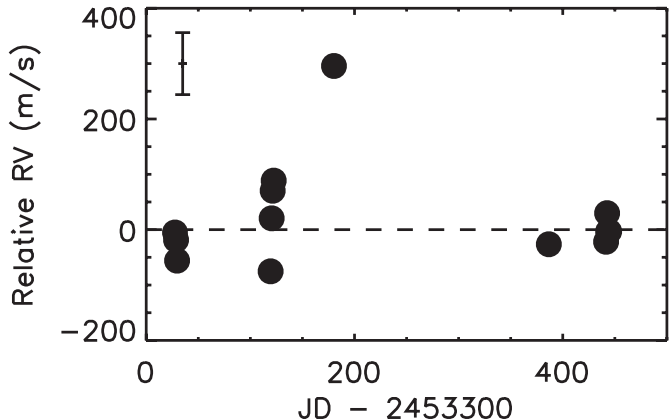}
\caption{RV curve of the slowly rotating young star TWA 7; an 
average errorbar is illustrated in the upper left.  All but 1 
of the RVs are consistent to within 100 m/s ($\sigma = 47$).  The 
one outlier is possibly due to a transient stellar event such as a 
flare.
\label{rv_twa7}}
\end{figure}

Altogether these results confirm the expectation that infrared 
observations mitigate the RV variations due to stellar activity 
and, moreover, quantify the benefit.  As discussed in the introduction,
this is expected if the RV variations are primarily caused by cool star
spots.  However, infrared observations made during other chromospheric 
events 
such as giant flares or coronal mass ejections may have little or no
benefit over optical measurements.  It has been shown that giant flares
on active stars can cause temporary shifts in optically determined RV of 
several hundred m/s \citep{reiners09}.  Although we do not have 
information to identify flare or flare-like events in our data, this
could explain the occasional large ($>$ 200 m/s), one-time RV changes
seen for some
stars.  One of our clearest examples occured during observations of
the slowly rotating star TWA 7 ($v$sin$i$ = 4.5 km/s).  As illustrated 
in Figure \ref{rv_twa7}, all but 
1 of its 12 RV measurements are consistent to within 100 m/s 
($\sigma_{obs}$ = 47 m/s); the discrepant epoch deviates by $\sim 300$
m/s from the mean RV of the other 11 epochs.  Comparisons of stars
observed immediately before and after this epoch show no similar
large amplitude deviations, which might be expected if the detector
was especially noisy at this time; the deviation appears to be limited 
to this observation.  The SNR of this observation is also above the
mean of the remaining observations, and the spectrum shows no 
defects, such as an cosmic ray event that was not corrected in the
optimal extraction.  This leads us to believe the deviation is 
intrinsic to the star itself.  Because we can't fully explain the 
event (or others like it), we highlight it here as a cautionary remark 
for future surveys.
Coordinated photometric and/or spectroscopic observations that 
target chromospheric indicators may help identify these events, 
which when excluded, would improve the overall sensitivity to low 
mass companions.

Finally, the ensemble infrared stellar jitter results provide 
a guide for the achievable RV precision for young active stars in 
the limit of 
high SNR spectra with no wavelength calibration error.  For 
the slowest rotating stars ($v$sin$i$ $< 6$ km/s), the median stellar 
jitter is 77 m/s (1 $\sigma$ spread of 21 m/s).  For stars with modest 
rotation (6 km/s $< v$sin$i <$ 12 km/s), the median stellar jitter is 
108 m/s (1 $\sigma$ spread of 43 m/s).  And for the most rapidly 
rotating stars ($v$sin$i$ of 12 - 23 km/s), the median stellar jitter 
is 168 m/s (1 $\sigma$ spread of 24 m/s).  We also note that none 
of the 4 stars with $v$sin$i$ values $> 12$ km/s have stellar jitter 
values above 180 m/s, possibly hinting at stellar jitter saturation
(Figure \ref{stellar_jitter}).  Although the available data cannot 
robustly confirm this, we highlight this possibility because this 
$v$sin$i$ limit is similar to the observed X-ray saturation 
limit of $\sim 15$ km/s, though for slightly more massive
stars in the Pleiades \citep{stauffer94}; the X-ray saturation is
interpreted as a consequence of maximal coverage of star spots.  
Likewise, we find that stars with the largest stellar jitter have 
rotational periods $< 3.3$ days (Table 1).  As demonstrated in 
\citep{pizzolato03}, subsolar mass stars with rotation periods 
$<$ 3.0 - 3.5 days (depending somewhat on mass) all have saturated 
X-ray emission.  Given that the most rapidly rotating stars 
studied here appear to be near or beyond the regime where X-rays 
saturate, its plausible to presume that the stellar jitter is 
also saturated.  The results have implications for how precisely
the radial velocities of rapidly rotating stars can be determined.

\section{Summary}
We present the results of a high-precision infrared RV study of 
20 young stars in the $\beta$ Pic and TW Hya Associations, as 
well as the chromospherically active, young disk M3.5 star GJ 873.  
High spectral resolution (R $\sim$ 30,000) 
measurements at 2.3 $\mu$m were obtained with NIRSPEC at the 
Keck Observatory; the majority of stars have more than a dozen 
epochs, with a typical temporal sampling of several observations 
over a week-long run, and then several runs over 2 to 3 years.
Precise RVs are determined using telluric absorption features as 
an absolute wavelength reference.  Each observation is modeled as 
the combination of a telluric spectrum and a synthetically generated 
stellar spectrum, both convolved by a parametrized instrumental 
profile and projected onto a parametrized wavelength solution. The 
best fit, which includes the RV of the star, is determined 
by minimizing the $\chi^2$.

This modeling technique yields RV dispersions of $\sim 50$ m/s for 
13+ epochs each of 3 slowly rotating field stars.  The dispersion of 
these RV measurements is dominated by instrumental errors and 
thus $\sim 50$ m/s can be considered a practical precision limit 
for NIRPSEC
\citep[see also][]{blake10}; this precision is nevertheless well above 
NIRSPEC's original design specifications.  Significant improvements
to this precision will likley require a new detector to mitigate 
dominant noise sources such as fringing, variable bias, and read 
noise.

The observed RV dispersions for young stars range from 
48 m/s to 197 m/s.  These dispersions are dominated by noise
from stellar activity, or stellar jitter.  We estimate the 
contribution from stellar jitter by subtracting, in quadrature, 
the average instrumental noise (46 m/s) and the calculated 
theoretical noise ($\sim 40$ m/s) from the observed dispersions.  
The stellar jitter increases with projected rotational velocity
($v$sin$i$).  The slowest rotating stars ($v$sin$i$ $< 6$ km/s)
have a median stellar jitter of 77 m/m, modest rotation stars 
(6 km/s $< v$sin$i <$ 12 km/s) have a median stellar jitter of
108 m/s, and the most rapidly rotating stars ($v$sin$i$ of 12 - 
23 km/s) have a median stellar jitter of 168 m/s.  There is
also tentative evidence that stellar jitter saturates at $\sim 
180$ m/s above $v$sin$i$ values of $\sim 12$ km/s, or below
rotational periods of $\sim 3-4$ days, consistent with the
rate of rotation that causes X-rays to saturate.

These infrared stellar jitter values are, on average, a factor
of 3 less than the stellar jitter values of similar age $\beta$ 
Pic Association stars observed at optical (R-band) wavelengths.  
Case studies  by \citet{huelamo08} and \citet{mahmud11} corroborate 
these reductions.  
This is expected if stellar jitter is predominantly caused by cool 
star spots, and confirms that infrared measurements have a 
significant advantage over optical measurements in obtaining 
precise RV measurements of active stars.

Three stars have RV dispersions significantly above the level 
expected for their $v$sin$i$.  The smoothly accelerating
RVs of 2 of these stars, GJ3305 and TWA 23, indicate they are 
single-lined spectroscopic binaries.  The RV orbits are still 
under-determined.  In the case of GJ 3305, the motion is likely 
caused by its recently identified $\sim 93$ milliarcsecond 
companion \citep{kasper07}.  The 3rd star, TWA 13A, exhibits 
large amplitude ($> 200$ m/s), short-timescale variations indicative 
of a hot Jupiter-like companion, but the available data are 
insufficient to confirm this.  We label it as a candidate RV 
variable.  We also note that the relative rarity of hot Jupiter
companions ($\sim 1\%$) implies that a first discovery in this 
small sample of 20 young stars would be most fortuitous.

For the remainder of the sample, these observations exclude the 
presence of any 'hot' (P $< 3$ days) companions more massive 
than 8 M$_{Jup}$, and any 'warm' (P $< 30$ day) companions 
more massive than 17 M$_{Jup}$, on average.  Assuming an 
edge-on orbit for the edge-on disk system AU Mic, these 
observations exclude the presence of any hot Jupiters more 
massive than 1.8 M$_{Jup}$ or warm Jupiters more massive 
than 3.9 M$_{Jup}$.  If massive Jupiters exist in the AU Mic 
system, they have not migrated inwards in its $\sim$ 12 Myr 
lifetime.

While the effect of star spots will continue to be a limiting
factor in the search for young short-period planets, even at 
infrared wavelengths, the results presented here elucidate 
better the opportunities for determining precise RVs from 
infrared observations in the limit of high stellar activity.




\acknowledgments

We are grateful to the support staff at Keck observatory, especially Grant 
Hill and Jim Lyke, and to the scientists who built and continue to maintain
NIRSPEC, especially Dr. Ian McLean.  We appreciate the data provided by the 
NASA/IPAC Infrared Science Archive and the privilege to observe on the 
revered summit of Mauna Kea.

\clearpage

%
%

\appendix
\section{Individual Radial Velocity Measurements}

Below are the individual RV measurements for the 24 stars observed in this 
survey.  The first column provides the star name, the number of RV 
measurements, the mean radial velocity its uncertainty, and the standard 
deviation of RV measurements.  The Julian Dates of the observation are 
listed in the 2nd column and the barycentric RVs and associated 
uncertainties are listed in the 3rd column.

\LongTables
\begin{deluxetable}{llll}
\tablecolumns{4} 
\tabletypesize{\scriptsize}
\tablecaption{Observational Sample \label{table3}}
\tablewidth{0pt}
\tablehead{
 \colhead{}
& \colhead{}
& \colhead{Radial Velocity}
& \colhead{} \\
\colhead{Star}
& \colhead{HJD - 2,400,000}
& \colhead{(m/s)}
& \colhead{SNR}
}
\startdata
GJ 628 & 53522.337 & -21,061 $\pm$ 51 & 141\\
  N=13 & 53523.502 & -21,077 $\pm$ 52 & 203\\
 $\bar{RV}=-21,113 \pm 15 $ & 53597.224 & -21,139 $\pm$ 53 & 183\\
$\sigma=54.84$ & 53741.688 & -21,104 $\pm$ 57 & 141\\
   & 53742.688 & -21,140 $\pm$ 52 & 198\\
   & 53743.685 & -21,145 $\pm$ 46 & 264\\
   & 53929.374 & -21,084 $\pm$ 53 & 221\\
  & 53930.307 & -21,028 $\pm$ 53 & 147\\
   & 53931.387 & -21,123 $\pm$ 53 & 205\\
   & 54308.353 & -21,251 $\pm$ 53 & 208\\
   & 54309.355 & -21,145 $\pm$ 49 & 278\\
  & 54311.343 & -21,095 $\pm$ 51 & 214\\
   & 54312.274 & -21,083 $\pm$ 51 & 241\\
 \hline
 GJ 725A & 53419.683 & -557 $\pm$ 57 & 161\\
  N=18 & 53420.687 & -587 $\pm$ 54 & 186\\
$\bar{RV}=-611 \pm 12 $ & 53421.580 & -553 $\pm$ 51 & 221\\
 $\sigma=51.24$ & 53422.638 & -647 $\pm$ 51 & 230\\
   & 53522.547 & -627 $\pm$ 53 & 187\\
   & 53523.419 & -546 $\pm$ 54 & 191\\
  & 53523.618 & -608 $\pm$ 50 & 261\\
   & 53596.363 & -543 $\pm$ 53 & 191\\
   & 53597.343 & -703 $\pm$ 50 & 248\\
   & 53670.202 & -707 $\pm$ 47 & 206\\
   & 53928.535 & -578 $\pm$ 50 & 275\\
  & 53929.453 & -631 $\pm$ 51 & 260\\
   & 53930.410 & -623 $\pm$ 59 & 147\\
   & 53931.463 & -614 $\pm$ 52 & 173\\
   & 54308.364 & -636 $\pm$ 54 & 198\\
   & 54309.399 & -561 $\pm$ 55 & 172\\
  & 54311.377 & -680 $\pm$ 54 & 176\\
   & 54312.291 & -605 $\pm$ 53 & 196\\
 \hline
 GJ 725B & 53419.682 & 1,305 $\pm$ 55 & 178\\
  N=18 & 53420.686 & 1,319 $\pm$ 59 & 144\\
$\bar{RV}=1,321 \pm 13 $ & 53421.578 & 1,352 $\pm$ 52 & 150\\
 $\sigma=53.07$ & 53422.636 & 1,329 $\pm$ 50 & 261\\
   & 53522.549 & 1,410 $\pm$ 52 & 194\\
   & 53523.421 & 1,411 $\pm$ 51 & 233\\
  & 53523.621 & 1,329 $\pm$ 52 & 233\\
   & 53596.365 & 1,376 $\pm$ 53 & 196\\
   & 53597.344 & 1,278 $\pm$ 54 & 183\\
   & 53670.204 & 1,388 $\pm$ 47 & 151\\
  & 53928.538 & 1,324 $\pm$ 52 & 241\\
   & 53929.455 & 1,266 $\pm$ 55 & 200\\
   & 53930.412 & 1,282 $\pm$ 54 & 194\\
   & 53931.489 & 1,347 $\pm$ 53 & 205\\
  & 54308.367 & 1,278 $\pm$ 57 & 172\\
   & 54309.401 & 1,226 $\pm$ 55 & 172\\
   & 54311.379 & 1,253 $\pm$ 53 & 206\\
   & 54312.293 & 1,304 $\pm$ 52 & 218\\
 \hline
GJ 182 & 53327.568 & 19,753 $\pm$ 53 & 172\\
  N=6 & 53328.596 & 19,732 $\pm$ 55 & 184\\
 $\bar{RV}=19,818 \pm 42 $ & 53686.442 & 19,696 $\pm$ 52 & 166\\
$\sigma=102.91$ & 53741.468 & 19,941 $\pm$ 50 & 224\\
   & 53742.409 & 19,898 $\pm$ 56 & 175\\
   & 53743.389 & 19,887 $\pm$ 57 & 176\\
 \hline
 GJ 873 & 53327.225 & 606 $\pm$ 62 & 142\\
 N=21 & 53328.198 & 656 $\pm$ 47 & 141\\
 $\bar{RV}=545 \pm 25 $ & 53522.592 & 502 $\pm$ 58 & 147\\
 $\sigma=112.93$ & 53523.587 & 418 $\pm$ 55 & 136\\
   & 53596.442 & 559 $\pm$ 55 & 175\\
  & 53597.401 & 547 $\pm$ 54 & 165\\
   & 53669.328 & 303 $\pm$ 56 & 136\\
   & 53670.352 & 442 $\pm$ 54 & 179\\
  & 53686.247 & 558 $\pm$ 52 & 198\\
   & 53742.205 & 710 $\pm$ 55 & 167\\
   & 53743.187 & 719 $\pm$ 58 & 150\\
   & 53744.187 & 780 $\pm$ 48 & 128\\
   & 53928.569 & 469 $\pm$ 57 & 179\\
  & 53929.516 & 525 $\pm$ 51 & 215\\
   & 53929.615 & 476 $\pm$ 48 & 383\\
   & 53930.538 & 385 $\pm$ 55 & 170\\
   & 53931.493 & 542 $\pm$ 49 & 160\\
   & 54308.411 & 589 $\pm$ 57 & 149\\
  & 54309.429 & 565 $\pm$ 53 & 152\\
   & 54311.430 & 554 $\pm$ 57 & 164\\
   & 54312.359 & 541 $\pm$ 52 & 201\\
 \hline
 AU Mic & 53522.559 & -4,083 $\pm$ 54 & 181\\
 N=14 & 53523.549 & -4,151 $\pm$ 54 & 193\\
 $\bar{RV}=-4,130 \pm 33 $ & 53596.369 & -4,020 $\pm$ 52 & 206\\
 $\sigma=124.74$ & 53597.379 & -4,112 $\pm$ 51 & 228\\
  & 53669.193 & -4,018 $\pm$ 52 & 205\\
   & 53670.196 & -3,869 $\pm$ 47 & 212\\
   & 53928.504 & -4,206 $\pm$ 49 & 336\\
   & 53929.448 & -4,280 $\pm$ 50 & 274\\
  & 53930.456 & -4,225 $\pm$ 54 & 187\\
   & 53931.396 & -4,156 $\pm$ 50 & 248\\
   & 54308.426 & -4,008 $\pm$ 53 & 211\\
   & 54309.411 & -4,298 $\pm$ 50 & 284\\
  & 54311.404 & -4,111 $\pm$ 54 & 195\\
   & 54312.356 & -4,289 $\pm$ 52 & 205\\
 \hline
 AG Tri A & 53327.386 & 6,692 $\pm$ 53 & 208\\
  N=14 & 53328.376 & 6,650 $\pm$ 57 & 163\\
$\bar{RV}=6,743 \pm 26 $ & 53329.442 & 6,650 $\pm$ 56 & 151\\
 $\sigma=97.53$ & 53421.202 & 6,780 $\pm$ 53 & 196\\
   & 53422.198 & 6,766 $\pm$ 53 & 198\\
   & 53669.334 & 6,759 $\pm$ 50 & 228\\
  & 53670.356 & 6,723 $\pm$ 51 & 226\\
   & 53686.257 & 6,552 $\pm$ 53 & 163\\
   & 53686.512 & 6,620 $\pm$ 57 & 140\\
   & 53742.259 & 6,830 $\pm$ 56 & 200\\
  & 53744.196 & 6,878 $\pm$ 48 & 149\\
   & 53928.604 & 6,830 $\pm$ 64 & 123\\
   & 53929.591 & 6,824 $\pm$ 52 & 223\\
   & 53930.583 & 6,840 $\pm$ 52 & 213\\
 \hline
AG Tri B & 53327.389 & 5,845 $\pm$ 51 & 260\\
  N=14 & 53328.378 & 5,939 $\pm$ 53 & 212\\
 $\bar{RV}=5,961 \pm 35 $ & 53329.445 & 6,124 $\pm$ 60 & 126\\
$\sigma=132.05$ & 53421.204 & 6,126 $\pm$ 54 & 193\\
   & 53422.201 & 5,920 $\pm$ 52 & 221\\
   & 53669.336 & 5,888 $\pm$ 50 & 252\\
   & 53670.359 & 5,699 $\pm$ 51 & 231\\
  & 53686.259 & 5,871 $\pm$ 52 & 192\\
   & 53686.514 & 5,843 $\pm$ 52 & 210\\
   & 53742.262 & 6,147 $\pm$ 56 & 240\\
   & 53744.198 & 6,073 $\pm$ 48 & 167\\
  & 53928.609 & 6,075 $\pm$ 68 & 116\\
   & 53929.597 & 5,981 $\pm$ 50 & 271\\
   & 53930.587 & 5,923 $\pm$ 52 & 227\\
 \hline
 GJ 3305 & 53328.592 & 19,862 $\pm$ 52 & 209\\
 N=5 & 53686.440 & 20,548 $\pm$ 58 & 159\\
 $\bar{RV}=20,625 \pm 204 $ & 53741.455 & 20,817 $\pm$ 52 & 174\\
 $\sigma=457.20$ & 53742.407 & 20,954 $\pm$ 51 & 242\\
  & 53743.387 & 20,946 $\pm$ 52 & 237\\
 \hline
 GJ 799 A & 53522.564 & -3,614 $\pm$ 55 & 175\\
  N=14 & 53523.539 & -3,670 $\pm$ 53 & 129\\
$\bar{RV}=-3,727 \pm 40 $ & 53596.372 & -3,902 $\pm$ 49 & 273\\
 $\sigma=151.32$ & 53597.381 & -4,055 $\pm$ 51 & 218\\
   & 53669.188 & -3,556 $\pm$ 57 & 151\\
  & 53670.193 & -3,572 $\pm$ 47 & 204\\
   & 53928.490 & -3,911 $\pm$ 50 & 207\\
   & 53929.439 & -3,722 $\pm$ 54 & 193\\
   & 53930.450 & -3,658 $\pm$ 52 & 164\\
  & 53931.391 & -3,621 $\pm$ 49 & 294\\
   & 54308.357 & -3,749 $\pm$ 53 & 203\\
   & 54309.406 & -3,850 $\pm$ 54 & 185\\
   & 54311.400 & -3,727 $\pm$ 53 & 196\\
  & 54312.351 & -3,570 $\pm$ 52 & 204\\
 \hline
 GJ 799 B & 53522.561 & -4,946 $\pm$ 55 & 160\\
  N=14 & 53523.547 & -5,358 $\pm$ 60 & 181\\
$\bar{RV}=-5,126 \pm 55 $ & 53596.374 & -5,165 $\pm$ 53 & 253\\
 $\sigma=206.52$ & 53597.383 & -4,726 $\pm$ 49 & 195\\
   & 53669.190 & -5,020 $\pm$ 57 & 195\\
  & 53670.190 & -4,888 $\pm$ 49 & 173\\
   & 53928.495 & -4,969 $\pm$ 52 & 235\\
   & 53929.445 & -5,071 $\pm$ 56 & 216\\
   & 53930.454 & -5,310 $\pm$ 59 & 181\\
  & 53931.394 & -5,198 $\pm$ 54 & 227\\
   & 54308.360 & -5,091 $\pm$ 58 & 193\\
   & 54309.408 & -5,326 $\pm$ 55 & 217\\
   & 54311.402 & -5,477 $\pm$ 60 & 179\\
  & 54312.353 & -5,217 $\pm$ 55 & 211\\
 \hline
 GJ 871.1 A & 53327.203 & 3,025 $\pm$ 80 & 151\\
  N=14 & 53328.183 & 3,082 $\pm$ 55 & 181\\
$\bar{RV}=3,087 \pm 36 $ & 53329.187 & 3,185 $\pm$ 55 & 107\\
 $\sigma=134.03$ & 53522.582 & 3,036 $\pm$ 51 & 279\\
   & 53523.604 & 2,831 $\pm$ 60 & 165\\
   & 53596.432 & 3,099 $\pm$ 51 & 290\\
  & 53597.394 & 3,270 $\pm$ 55 & 202\\
   & 53669.269 & 3,054 $\pm$ 53 & 250\\
   & 53686.232 & 3,294 $\pm$ 49 & 146\\
   & 53928.518 & 3,103 $\pm$ 56 & 162\\
  & 53929.523 & 3,266 $\pm$ 57 & 209\\
   & 53930.529 & 3,090 $\pm$ 56 & 205\\
   & 54311.415 & 2,953 $\pm$ 52 & 267\\
   & 54312.414 & 2,931 $\pm$ 51 & 274\\
 \hline
GJ 871.1 B & 53327.211 & 2,151 $\pm$ 63 & 244\\
  N=15 & 53328.188 & 2,031 $\pm$ 65 & 229\\
 $\bar{RV}=2,031 \pm 44 $ & 53329.195 & 2,311 $\pm$ 76 & 170\\
$\sigma=168.83$ & 53522.587 & 1,872 $\pm$ 62 & 228\\
   & 53523.609 & 1,834 $\pm$ 71 & 187\\
   & 53596.436 & 2,125 $\pm$ 65 & 203\\
   & 53597.397 & 2,017 $\pm$ 68 & 189\\
  & 53669.272 & 1,722 $\pm$ 79 & 151\\
   & 53686.236 & 2,298 $\pm$ 50 & 181\\
   & 53928.525 & 2,069 $\pm$ 61 & 242\\
   & 53930.532 & 2,164 $\pm$ 66 & 204\\
  & 54308.428 & 2,056 $\pm$ 77 & 160\\
   & 54309.422 & 2,053 $\pm$ 58 & 277\\
   & 54311.419 & 1,923 $\pm$ 59 & 264\\
   & 54312.418 & 1,846 $\pm$ 58 & 277\\
 \hline
HIP 12545 & 53328.384 & 8,407 $\pm$ 54 & 219\\
  N=14 & 53329.401 & 8,399 $\pm$ 73 & 100\\
 $\bar{RV}=8,253 \pm 48 $ & 53421.208 & 8,422 $\pm$ 60 & 152\\
$\sigma=179.25$ & 53422.205 & 8,201 $\pm$ 54 & 202\\
   & 53669.402 & 7,945 $\pm$ 55 & 193\\
   & 53670.393 & 8,145 $\pm$ 54 & 195\\
   & 53686.263 & 8,219 $\pm$ 55 & 166\\
  & 53686.519 & 8,077 $\pm$ 54 & 182\\
  & 53742.255 & 8,438 $\pm$ 62 & 160\\
   & 53744.203 & 8,495 $\pm$ 48 & 140\\
   & 53928.614 & 7,985 $\pm$ 65 & 129\\
   & 53929.608 & 8,116 $\pm$ 50 & 295\\
  & 53930.592 & 8,325 $\pm$ 53 & 211\\
   & 53931.641 & 8,372 $\pm$ 55 & 213\\
 \hline
 TWA 7 & 53327.632 & 12,419 $\pm$ 51 & 197\\
  N=13 & 53328.630 & 12,406 $\pm$ 47 & 183\\
$\bar{RV}=12,446 \pm 26 $ & 53329.631 & 12,368 $\pm$ 57 & 121\\
 $\sigma=94.18$ & 53419.453 & 12,349 $\pm$ 52 & 195\\
   & 53420.469 & 12,445 $\pm$ 48 & 291\\
  & 53421.436 & 12,495 $\pm$ 50 & 232\\
   & 53422.442 & 12,513 $\pm$ 51 & 204\\
   & 53480.294 & 12,720 $\pm$ 50 & 220\\
   & 53686.641 & 12,398 $\pm$ 52 & 158\\
  & 53741.575 & 12,402 $\pm$ 52 & 180\\
   & 53742.546 & 12,454 $\pm$ 52 & 187\\
   & 53743.549 & 12,421 $\pm$ 51 & 188\\
   & 53744.523 & 12,403 $\pm$ 48 & 220\\
 \hline
TWA 8A & 53327.636 & 8,610 $\pm$ 52 & 175\\
  N=15 & 53328.653 & 8,688 $\pm$ 51 & 192\\
 $\bar{RV}=8,679 \pm 23 $ & 53329.635 & 8,705 $\pm$ 57 & 124\\
$\sigma=90.07$ & 53419.473 & 8,555 $\pm$ 50 & 243\\
   & 53420.484 & 8,542 $\pm$ 52 & 183\\
   & 53421.460 & 8,667 $\pm$ 50 & 225\\
   & 53422.462 & 8,647 $\pm$ 48 & 208\\
  & 53480.306 & 8,877 $\pm$ 51 & 217\\
   & 53522.258 & 8,599 $\pm$ 49 & 186\\
   & 53523.248 & 8,642 $\pm$ 51 & 214\\
   & 53686.645 & 8,701 $\pm$ 50 & 195\\
  & 53741.605 & 8,822 $\pm$ 50 & 231\\
   & 53742.590 & 8,729 $\pm$ 50 & 252\\
   & 53743.562 & 8,672 $\pm$ 49 & 242\\
   & 53744.548 & 8,731 $\pm$ 49 & 257\\
 \hline
TWA 8B & 53327.640 & 8,489 $\pm$ 62 & 130\\
  N=15 & 53328.650 & 8,523 $\pm$ 63 & 126\\
 $\bar{RV}=8,607 \pm 32 $ & 53329.639 & 8,453 $\pm$ 63 & 122\\
$\sigma=123.46$ & 53419.480 & 8,572 $\pm$ 57 & 170\\
   & 53420.487 & 8,537 $\pm$ 57 & 162\\
   & 53421.462 & 8,575 $\pm$ 60 & 140\\
   & 53422.465 & 8,523 $\pm$ 51 & 170\\
  & 53480.310 & 8,813 $\pm$ 54 & 197\\
   & 53522.266 & 8,537 $\pm$ 58 & 164\\
   & 53523.252 & 8,477 $\pm$ 53 & 219\\
   & 53686.649 & 8,633 $\pm$ 55 & 164\\
  & 53741.610 & 8,718 $\pm$ 54 & 202\\
   & 53742.594 & 8,761 $\pm$ 55 & 194\\
   & 53743.567 & 8,662 $\pm$ 54 & 197\\
   & 53744.555 & 8,833 $\pm$ 51 & 253\\
 \hline
TWA 9A & 53419.487 & 11,569 $\pm$ 55 & 181\\
  N=11 & 53420.493 & 11,633 $\pm$ 54 & 191\\
 $\bar{RV}=11,649 \pm 21 $ & 53421.466 & 11,574 $\pm$ 53 & 155\\
$\sigma=70.61$ & 53422.469 & 11,567 $\pm$ 57 & 159\\
   & 53480.315 & 11,716 $\pm$ 54 & 184\\
   & 53523.227 & 11,659 $\pm$ 55 & 173\\
   & 53741.616 & 11,750 $\pm$ 55 & 175\\
  & 53742.600 & 11,569 $\pm$ 57 & 164\\
   & 53743.572 & 11,686 $\pm$ 57 & 179\\
   & 53744.577 & 11,734 $\pm$ 52 & 215\\
   & 54962.239 & 11,680 $\pm$ 54 & 198\\
 \hline
TWA 9B & 53419.492 & 12,271 $\pm$ 58 & 150\\
  N=11 & 53420.497 & 12,320 $\pm$ 54 & 175\\
$\bar{RV}=12,279 \pm 27 $ & 53421.470 & 12,101 $\pm$ 58 & 144\\
 $\sigma=90.25$ & 53422.472 & 12,199 $\pm$ 56 & 160\\
   & 53480.319 & 12,383 $\pm$ 54 & 176\\
  & 53523.232 & 12,367 $\pm$ 56 & 168\\
   & 53741.621 & 12,335 $\pm$ 57 & 156\\
   & 53742.605 & 12,282 $\pm$ 53 & 200\\
   & 53743.577 & 12,186 $\pm$ 52 & 201\\
  & 53744.584 & 12,247 $\pm$ 51 & 229\\
   & 54962.245 & 12,380 $\pm$ 51 & 226\\
 \hline
 TWA 11B & 53420.507 & 9,073 $\pm$ 58 & 156\\
  N=11 & 53421.444 & 8,782 $\pm$ 55 & 191\\
$\bar{RV}=8,923 \pm 58 $ & 53421.487 & 8,719 $\pm$ 55 & 184\\
 $\sigma=190.79$ & 53422.455 & 8,791 $\pm$ 53 & 153\\
   & 53480.334 & 8,564 $\pm$ 54 & 194\\
   & 53523.243 & 9,177 $\pm$ 57 & 171\\
  & 53741.633 & 8,992 $\pm$ 55 & 183\\
   & 53742.618 & 9,092 $\pm$ 56 & 186\\
   & 53743.592 & 8,861 $\pm$ 55 & 196\\
   & 53744.634 & 9,056 $\pm$ 47 & 198\\
  & 54963.240 & 9,043 $\pm$ 53 & 219\\
 \hline
 TWA 12 & 53328.641 & 12,276 $\pm$ 54 & 255\\
  N=12 & 53329.646 & 12,456 $\pm$ 64 & 163\\
$\bar{RV}=12,498 \pm 52 $ & 53419.457 & 12,352 $\pm$ 61 & 188\\
 $\sigma=180.79$ & 53420.479 & 12,708 $\pm$ 57 & 209\\
   & 53421.450 & 12,420 $\pm$ 56 & 228\\
  & 53422.459 & 12,445 $\pm$ 60 & 188\\
   & 53480.297 & 12,302 $\pm$ 60 & 186\\
   & 53686.664 & 12,445 $\pm$ 58 & 196\\
   & 53741.592 & 12,370 $\pm$ 60 & 188\\
  & 53742.557 & 12,655 $\pm$ 60 & 196\\
   & 53743.558 & 12,758 $\pm$ 59 & 192\\
   & 53744.530 & 12,789 $\pm$ 62 & 190\\
 \hline
 TWA 13A & 53327.646 & 11,889 $\pm$ 57 & 158\\
 N=14 & 53328.670 & 12,220 $\pm$ 55 & 181\\
 $\bar{RV}=11,668 \pm 65 $ & 53419.463 & 11,817 $\pm$ 53 & 218\\
 $\sigma=241.66$ & 53420.472 & 11,488 $\pm$ 56 & 172\\
  & 53421.454 & 11,469 $\pm$ 55 & 181\\
   & 53422.445 & 11,523 $\pm$ 56 & 175\\
   & 53480.303 & 11,559 $\pm$ 54 & 190\\
   & 53686.656 & 11,616 $\pm$ 55 & 171\\
  & 53741.598 & 11,365 $\pm$ 58 & 163\\
   & 53742.549 & 11,324 $\pm$ 56 & 187\\
   & 53743.551 & 11,901 $\pm$ 55 & 200\\
   & 53744.537 & 11,735 $\pm$ 53 & 226\\
  & 54962.223 & 11,752 $\pm$ 49 & 176\\
   & 54963.225 & 11,696 $\pm$ 50 & 211\\
 \hline
 TWA 13B & 53327.650 & 12,035 $\pm$ 56 & 167\\
  N=13 & 53328.673 & 11,677 $\pm$ 58 & 150\\
$\bar{RV}=12,075 \pm 41 $ & 53419.468 & 12,130 $\pm$ 53 & 209\\
 $\sigma=149.06$ & 53420.475 & 11,984 $\pm$ 53 & 214\\
   & 53421.457 & 12,071 $\pm$ 54 & 186\\
  & 53422.448 & 12,085 $\pm$ 55 & 182\\
   & 53480.301 & 12,302 $\pm$ 54 & 190\\
   & 53686.659 & 12,020 $\pm$ 54 & 179\\
   & 53741.601 & 12,260 $\pm$ 53 & 206\\
  & 53742.552 & 12,111 $\pm$ 56 & 206\\
   & 53743.554 & 12,050 $\pm$ 54 & 194\\
   & 53744.542 & 12,134 $\pm$ 53 & 217\\
   & 54962.235 & 12,109 $\pm$ 53 & 217\\
 \hline
TWA 23 & 53419.500 & 4,445 $\pm$ 61 & 244\\
  N=14 & 53420.503 & 4,504 $\pm$ 62 & 230\\
 $\bar{RV}=6,520 \pm 648 $ & 53421.439 & 4,741 $\pm$ 65 & 204\\
$\sigma=2425.03$ & 53421.483 & 4,573 $\pm$ 63 & 221\\
   & 53422.451 & 4,778 $\pm$ 53 & 243\\
   & 53480.324 & 7,155 $\pm$ 58 & 225\\
   & 53523.238 & 8,271 $\pm$ 56 & 244\\
  & 53686.669 & 9,022 $\pm$ 62 & 196\\
   & 53741.628 & 9,151 $\pm$ 62 & 207\\
   & 53742.613 & 9,242 $\pm$ 62 & 206\\
   & 53743.585 & 9,204 $\pm$ 60 & 224\\
  & 53744.596 & 9,298 $\pm$ 52 & 262\\
   & 54962.254 & 3,085 $\pm$ 60 & 253\\
   & 54963.233 & 3,816 $\pm$ 63 & 231\\
 \hline
\enddata
\tablecomments{Reported values are Heliocentric radial velocities.}
\end{deluxetable}


\begin{thebibliography}{}
\bibitem[Abranin et al. (1998)]{abranin98} Abranin, E. P., et al. 1998, A\&ApT, 17, 221
\bibitem[Adams \& Laughlin (2003)]{adamslaughlin03} Adams, F. C. \& Laughlin, G. 2003, 
2003, 163, 290
\bibitem[Alibert et al. (2005)]{alibert05} Alibert, Y., Mordasini, C., Benz, W. \& 
Winisdoerffer, C. \aap, 434, 343
\bibitem[Baliunas et al. (1995)]{baliunas95} Baliunas, S. L., et al. 1995, \apj, 438, 269
\bibitem[Baraffe et al. (1998)]{baraffe98} Baraffe, I., Charbrier, G., Allard, F. \& 
Hauschildt, P. H. 1998, \aap, 337, 403
\bibitem[Bean et al. (2010)]{bean10} Bean, J. L., Seifahrt, A., Hartmann, H., 
Nilsson, H., Wiedemann, G., Reiners, A., Dreizler, S. \& Henry, T. J. 2010, 
\apj, 713, 410
\bibitem[Blake et al. (2007)]{blake07} Blake, C., Charbonneau, D., White, R. J., 
Marley, M. S. \& Saumon, D. 2007, \apj, 666, 1198
\bibitem[Blake et al. (2010)]{blake10} Blake, C., Charbonneau, D. \& White, R. J. 2010, 
\apj, 723, 684
\bibitem[Boss (1997)]{boss97} Boss, A. P. 1997, Science, 276, 1836
\bibitem[Boss (2004)]{boss04} Boss, A. P. 2004, \apj, 610, 456
\bibitem[Boss (2006)]{boss06} Boss, A. P. 2006, \apjl, 644, 79
\bibitem[Brice\~no et al. (2001)] {briceno01} Brice\~no, C., et al. 2001, Science, 291, 
93
\bibitem[Butler et al. (2006)]{butler06} Butler, R. P. et al. 2006, \pasp, 118, 
1685
\bibitem[Butler et al. (1996)]{butler96} Butler, R. P., Marcy, G. W., Williams, E., 
McCarthy, C., Dosanjh, P. \& Vogt, S. S. 1996, \pasp, 108, 500 
\bibitem[Carpenter et al. (2006)]{carpenter06} Carpenter, J. M., Mamajek, E. E., 
Hillenbrand, L. A., Meyer, M. R. 2006, \apjl, 651, 49
\bibitem[Cochran et al. (1988)]{cochran88} Cochran, W. D. 1988, \apj, 334, 349
\bibitem[Cochran et al. (1997)]{cochran97} Cochran, W. D., Hatzes, A. P., Butler, R. P.
\& Marcy, G. W. 1997, ApJ, 483, 457
\bibitem[Crockett et al. (2011)]{crockett11} Crockett, C. J., Mahmud, N. I., Prato, L,
Johns-Krull, C. M., Jaffe, D. T. \& Beichman, C. A. 2011, \apj, in press
\bibitem[Cumming et al. (2008)]{cumming08} Cumming, A., Butler, R. P., Marcy, G. W., 
Vogt, S. S., Wright, J. T. \& Fischer, D. A. 2008, \pasp, 120, 531
\bibitem[Deming et al. (2005)]{deming05} Deming, D., Brown, T. M., Charbonneau, D., 
Harrington, J. \& Richardson, L. J. 2005, \apj, 622, 1149
\bibitem[Endl et al. (2006)]{endl06} Endl, M., Cochran, W. D., Tull, R. G. \& MacQueen, 
P. J. 2006, \aj, 132, 2755
\bibitem[Figueira et al. (2010)]{figueira10} Figueira, P., Marmier, M., Bonfils, X., 
di Folco, E., Udry, S., Santos, N. C., Lovis, C., M\'egevand, D., Melo, C. H. F. \& 
Pepe, F. 2010, \aap, 513, 8
\bibitem[Fischer, D. A. et al. (2006)]{fischer06} Fischer, D. A., et al. 2006, \apj,
637, 1094
\bibitem[Feigelson et al. (2006)]{feigelson06} Feigelson, E. D., Lawson, W. A., 
Stark, M., Townsley, L. \& Garmire, G. P.
\bibitem[Goldreich \& Tremaine (1980)]{goldreichtremaine80} Goldreich, P. \& 
Tremaine, S. 1980, \apj, 241, 425
\bibitem[Haisch (2001)]{haisch01} Haisch, K., Lada, E. A. \& Lada, C. J. 2001, 
\apj, 553, 153
\bibitem[Hauschildt et al. (1999)]{hauschildt99} Hauschildt, P. H., Allard, F. \& 
Baron, E. 1999, Ap, 512, 377
\bibitem[Hawley et al. (1996)]{hawley96} Hawley, S. L., Gizis, J. E. \& Reid, I., N. 
1996, \aj, 112, 2799
\bibitem[Hebb et al. (2007)]{hebb07} Hebb, L. et al. 2007, \mnras, 379, 63
\bibitem[Henry et al. (1993)]{henry94} Henry, T. J., Kirkpatrick, J. D. \&  
Simons, D. A.
1994, \aj, 108, 1437
\bibitem[Henry \& McCarthy (1993)]{henrymccarthy93} Henry, T. J. \& McCarthy, D. W., Jr.
1993, \aj, 106, 773
\bibitem[Herbst et al. (1994)]{herbst94} Herbst, W., Herbst, D. K., Grossman, E. J., 
Weinstein, D. 1994, \aj, 108, 1906
\bibitem[Hern\'an-Obispo et al. (2010)]{hernanobispo10} Hern\'an-Obispo, M., 
G\'alvez-Ortiz, M. C., Anglada-Escud\'e, G., Kane, S. R., Barnes, J. R., 
de Castro, E. \& Cornide, M. 2010, \aap, 512, 45
\bibitem[Hillenbrand \& White (2004)]{hillenbrandwhite04} Hillenbrand, L. A. \& White,
R. J. 2004, \apj, 604, 741
\bibitem[Hu\'elamo et al. (2008)]{huelamo08} Hu\'elamo, N., et al. 2008, \aap, 489, 9
\bibitem[Huerta et al. (2008)]{huerta08} Huerta, M., Johns-Krull, C. M., Prato, L., 
Hartigan, P. \& Jaffe, D. T. 2008, \apj, 678, 472
\bibitem[H\"unsch et al. (1999)]{hunsch99} H\"unsch, M., Schmitt, J. H. M. M., Sterzik, 
M. F. \& Voges, W. 1999, \aaps, 135, 319
\bibitem[Horne et al. (1986)]{horne86} Horne, K. 1986, \pasp, 98, 609
\bibitem[Howard et al. (2010)]{howard10} Howard, A. W., Marcy, G. W., Johnson, J. A., 
Fischer, D. A., Wright, J. T., Howard, I., Valenti, J. A., Anderson, J., Lin, 
D. N. C. \& Shigeru, I. 2010, Science, 330, 653
\bibitem[Ireland et al. (2011)]{ireland11} Ireland, M. J., Kraus, A., Martinache, F., 
Law, N. \& Hillenbrand, L. A. 2011, \apj, 726, 113
\bibitem[Johns-Krull \& Valenti (1996)]{johnskrullvalenti96} Johns-Krull, C. M. \& 
Valenti, J. A. 1996, \apj, 459, L95
\bibitem[Jones et al. (2006)]{jones06} Jones, H. R. A., Butler, R. P., Tinney, C. 
G., Marcy, G. W., Carter, B. D., Penny, A. J., McCarthy, C. \& Bailey, J. 2006,
\mnras, 369, 249
\bibitem[Lagrange et al. (2010)]{lagrange10} Lagrange, A. -M., Bonnefoy, M., Chauvin, 
G., Apai, D., Ehrenreich, D., 
Boccaletti, A., Gratadour, D., Rouan, D., Mouillet, D., Lacour, S. \& Kasper, M.
2010, Sci, 329, 57
\bibitem[Latham et al. (1989)]{latham89} Latham, D. W., Stefanik, R. P., Mazeh, T., 
Mayor, M. \& Burki, G. 1989, Natur, 339, 38
\bibitem[Livingston \& Wallace (1991)] {livingstonwallace91} Livingston, W., \& 
Wallace, L. 1991, Atlas of the Solar Spectrum in the Infrared from 1850 to 9000 
cm$^{-1}$ (NSO Tech. Rep.; Tucson: NSO)
\bibitem[Leggett (1992)]{leggett92} Leggett, S. 1992, \apjs, 82, 351
\bibitem[Lubow \& Ida (2010)]{lubowida10} Lubow, S. H. \& Ida, S. 2010, in 
\textit{Exoplanets}, edited by S. Seager, Tucson, AZ: University of Arizona Press,
p. 347-371
\bibitem[Kalas et al. (2004)]{kalas04} Kalas, P., Liu, M. C., Matthews, B. C. 2004,
Science, 303, 1990
\bibitem[Kalas et al. (2008)]{kalas08} Kalas, P., Graham, J. R., Chiang, E., Fitzgerald, 
M. P., Clampin, M., Kite, E. S., Stapelfeldt, K., Marois, C. \& Krist, J. 2008, 
Science, 322, 1345
\bibitem[Kasper et al. (2007)]{kasper07} Kasper, M., Apai, D., Janson, M \& Brandner, 
W. 2007, \aap, 472, 321
\bibitem[Kataria \& Simon (2010)]{katariasimon10} Kataria, T. \& Simon, M. 2010, \aj, 
140, 206
\bibitem[Kraus \& Hillenbrand (2007)]{kraushillenbrand07} Kraus, A. L. \& Hillenbrand, 
L. A. 2007, \aj, 134, 2340
\bibitem[Nidever et al. (2002)]{nidever02} Nidever, D. L., Marcy, G. W., Butler, R. P.,
Fischer, D. A., \& Vogt, S. S. 2002, \apjs, 141, 503
\bibitem[Kennedy \& Kenyon (2008)]{kennedykenyon08} Kennedy, G. M. \& Kenyon, S. J. 
2008, \apj, 673, 502
\bibitem[Lin et al. (1996)]{lin96} Lin, D. N. C., Bodenheimer, P. \& Richardson,
D. C. 1996, Nature, 380, 606
\bibitem[Luhman (2003)]{luhman03} Luhman, K. 2003, \apj, 593, 1093
\bibitem[Macintosh et al. (2008)]{macintosh08} Macintosh, B. A., et al. 2008, SPIE,
7015, 180
\bibitem[Mahmud et al. (2011)]{mahmud11} Mahmud, N., Crockett, C. J., Johns-Krull, 
C. M., Prato, L., Hartigan, P. M., Jaffe, D. T. \& Beichman, C. A. 2011, \apj,
in press.
\bibitem[Malmberg \& Davies (2009)]{malmbergdavies09} Malmberg, D. \& Davies, M. B. 
2009, \mnras, 394, 26
\bibitem[Malmberg et al. (2007)]{malmberg07} Malmberg, D., de Angeli, F., Davies, M. 
B., Church, R. P., Mackey, D. \& Wilkinson, M. I. 2007, \mnras, 378, 1207
\bibitem[Marchi et al. (2009)]{marchi09} Marchi, S., Ortolani, S., Nagasawa, M. \& 
Ida, S. 2009, \mnras, 394, 93
\bibitem[Marcy \& Butler (1996)]{marcybutler96} Marcy, G. W. \& Butler, R. P. 1996, 
\apj, 464, 147
\bibitem[Marois et al. (2008)]{marois08} Marois, C., Macintosh, B., Barman, T., 
Zuckerman, B., Song, I., Patience, J., Lafreni\'ere, D. \& Doyon, R. 2008, Science, 
322, 1348
\bibitem[Mathieu et al. (2007)]{mathieu07} Mathieu, R. D., Baraffe, I., Simon, M., 
Stassun, K. G. \& White, R. J. 2007, Protostars and Planets V, B. Reipurth, D. Jewitt,
and K. Keil (eds.), University of Arizona Press, Tucson, 951 pp., 2007., p.411-425
\bibitem[Mayer et al. (2002)]{mayer02} Mayer, L., Quinn, T., Wadsley, J. \& Stadel, J.
2002, Science, 298, 1756
\bibitem[Mayor \& Queloz (1995)]{mayorqueloz95} Mayor, M. \& Queloz, D. 1995, 
Nature, 378, 355
\bibitem[Meru \& Bate (2010)]{merubate10} Meru, F. \& Bate, M. R. 2010, \mnras, 406, 
2279
\bibitem[Meru \& Bate (2011)]{merubate11} Meru, F. \& Bate, M. R. 2011, \mnras, 411, 
1
\bibitem[Messina et al. (2010)]{messina10} Messina, S., Desidera, S., Turatto, M., 
Lanzafame, A. C. \& Guinan, E. F. 2010, \aap, 520, 15
\bibitem[McLean et al. (1998)]{mclean98} McLean, I. S., et al. 1998, Proc. SPIE, 
3354, 566
\bibitem[Mentuch et al. (2008)]{mentuch08} Mentuch, E., Brandeker, A., van Kerkwijk, M. H.,
Jayawardhana, R. \& Hauschildt, P. H. 2008, \apj, 689, 1127
\bibitem[Mizuno et al. (1980)]{mizuno80} Mizuno, H. 1980, PThPh, 64, 544
\bibitem[Muirhead et al. (2011)]{muirhead11} Muirhead, P. S., et al. 2011, \pasp, in press.
\bibitem[Nelder \& Mead (1965)]{neldermead65} Nelder, J. A. \& Mead, R. 1965, 
Computer Journal, 7, 308
\bibitem[Neuh\"auser et al. (2005)]{neuhauser05} Neuh\"auswer, R., Guenther, E. W., 
Wuchterl, G., Maugrauer, M., Bedalov, A. \& Hauschildt, P. H. 2005, \aap, 
435, 13
\bibitem[Osten et al. (2006)]{osten06} Osten, R. A., Hawley, S. L., Allred, J., 
Johns-Krull, C. M., Brown, A. \& Harper, G. M. 2006, \apj, 647, 1349
\bibitem[Pascucci et al. (2006)]{pascucci06} Pascucci, I., et al. 2006, \apj, 651, 
1177
\bibitem[Paulson \& Yelda (2006)]{paulsonyelda06} Paulson, D. B. \& Yelda, S. 
2006, \pasp, 118, 706
\bibitem[Pettersen et al. (1984)]{pettersen84} Pettersen, B. R., Evans, D. S., 
Coleman, L. A. 1984, 282, 214
\bibitem[Piskunov \& Valenti (2002)]{piskunovvalenti02} Piskunov, N., E. \& Valenti, J. A.
2002, \aap, 385, 1095
\bibitem[Pizzolato et al. (2003)]{pizzolato03} Pizzolato, N., Maggio, A., Micela, G., 
Sciortino, S. \& Ventura, P. 2003, \aap, 397, 147
\bibitem[Pollack et al. (1996)]{pollack96} Pollack, J. B., Hubickyj, O., Bodenheimer,
P., Lissauer, J. J., Podolak, M. \& Greenzweig, Y. 1996, Icarus, 124, 62
\bibitem[Pont et al. (2009)]{pont09} Pont, F., et al. 2009, \aap, 502, 695
\bibitem[Prato et al. (2008)]{prato08} Prato, L., Huerta, M., Johns-Krull, C. M., 
Mahmud, N., Jaffe, D. T. \& Hartigan, P. 2008, \apj, 687, 103
\bibitem[Press et al. (1992)]{press92} Press, W. H., Teulolsky, S. A., 
Vetterling, W. T. \& Flannery, B. P. 1992, Numerical Recipes in C. The Art of 
Scientific Computing, Cambridge: University Press, 1992, 2nd ed.
\bibitem[Queloz et al. (2001)]{queloz01} Queloz, D., Henry, G. W., Sivan, J. P., 
Baliunas, S. L., Beuzit, J. L., Donahue, R. A., Mayor, M., Naef, D., Perrier, C. \&
Udry, S. 2001, \aap, 379, 279
\bibitem[Raymond et al. (2011)]{raymond11} Raymond, S. N., et al. 2011, \aap, 530, 62
\bibitem[Reiners \& Basri (2007)]{reinersbasri07} Reiners, A. \& Basri, G. 2007, \aj, 
656, 1121
\bibitem[Reiners (2009)] {reiners09} Reiners, A. 2009, \aap, 498, 853
\bibitem[Reiners et al. (2010)] {reiners10} Reiners, A., Bean, J. L., Huber, K. F., 
Dreizler, S., Seifahrt, A. \& Czesla, S. 2010, \apj, 710, 432
\bibitem[Robinson et al. (2006)]{robinson06} Robinson, S. E., Laughlin, G., 
Bodenheimer, P. \& Fischer, D. 2006, \apj, 643, 484
\bibitem[Saar \& Donahue (1997)]{saardonahue97} Saar, S. H. \& Donahue, R. A. 1997, 
\apj, 485, 319
\bibitem[Sato et al. (2005)]{sato05} Sato, B., et al. 2005, \apj, 633, 465
\bibitem[Santos et al. (2005)]{santos05} Santos, N. C., Benz, W. \& Mayor, M. 2005,
Science, 310, 251
\bibitem[Seifahrt et al. (2008)]{seifahrtkaufl08} Seifahrt, A. \& K\"aufl, H. U. 2008, \aap,
491, 929
\bibitem[Seifahrt et al. (2010)]{seifahrt10} Seifahrt, A., K\"aufl, H. U., Z\"angl, G.,
Bean, J. L., Richter, M. J. \& Siebenmorgen, R. 2010, \aap, 524, 11
\bibitem[Setiawan et al. (2008)]{setiawan08} Setiawan, J., Henning, Th., Launhardt, R.
M\"uller, A., Weise, P., K\"urster, M. 2008, Nature, 451, 38
\bibitem[Setiawan et al. (2007)]{setiawan07} Setiawan, J., Weise, P., Henning, Th., 
Launhardt, R., M\"uller, A. \& Rodmann, J.
\bibitem[Siess et al. (2000)]{siess00} Siess, L., Dufour, E. \& Forestini, M. 2000, 
\aap, 358, 593
\bibitem[Song et al. (2002)]{song02} Song, I., Zuckerman, B. \& Bessell, M. S. 2003, 
\apj, 599, 342
\bibitem[Song et al. (2003)]{song03} Song, I., Bessell, M. S. \& Zuckerman, B. 2003, 
\apj, 581, 43
\bibitem[Stauffer et al. (1994)]{stauffer94} Stauffer, J. R., Caillault, J.-P., 
Gagn\'e, M., Prosser, C. F. \& Hartmann, L. W. 1994, \apjs, 91, 625
\bibitem[Sterzik et al. (1999)]{sterzik96} Sterzik, M. F., Alcal\'a, J. M., Covino, E.
\& Petr, M. G. 1999, \aap, 346, 41
\bibitem[Tamuz et al. (2008)]{tamuz08} Tamuz, O. et al. 2008, \aap, 480, 33
\bibitem[Torres et al. (2008)]{torres08} Torres, c. A. O., Quast, G. R., Melo, C.
H. F. \& Sterzik, M. F. 2008, in 'Handbook of Star Forming Regions, Volume II: 
The Southern Sky', p. 757, ASP Monograph Publications, ed. B. Reipurth,
\bibitem[Udry et al. (2007)]{udrysantos07} Udry, S. \& Santos, N., C. 2007, \araa, 
45, 397
\bibitem[Walker et al. (1995)]{walker95} Walker, G. A. H., Walker, A. R., Irwin, 
A. W., Larson, A. M., Yang, S. L. S. \& Richardson, D. C. 1995, Icarus, 116, 359
\bibitem[Webb et al. (1999)]{webb99} Webb, R. A., Zuckerman, B., Platais, I., 
Patience, J., White, R. J., Schwartz, M., J. \& McCarthy, C. 1999, \apj, 512, 63
\bibitem[Wright et al. (2010)]{wright10} Wright, J. T., Fakhouri, O., Marcy, G. W., 
Eunkyu, H., Ying, F., Johnson, J. A., Howard, A. W., Fischer, D. A., Valenti, J. A.,
Anderson, J. \& Piskunov, N. 2010, \pasp, in print
\bibitem[Zuckerman \& Song (2004)]{zuckermansong04} Zuckerman, B. \& Song, I. 2004, 
\araa, 42, 685
\end{thebibliography}
\end{document}